%% file: Konda.Hofman.ea.ECC22.tex
\documentclass[conference]{ieeeconf}
\renewcommand{\baselinestretch}{0.995}

\usepackage{cite}
\usepackage{array}

\usepackage{amsthm}
\usepackage{amssymb}
\usepackage{lettrine}
\newcounter{thm}
\newtheorem{prob}[thm]{Problem}

\usepackage[pdftex]{graphicx}
\DeclareGraphicsExtensions{.pdf,.jpeg,.png,.eps}
\usepackage{url}
\usepackage{booktabs}
\usepackage{lettrine}
\usepackage{tikz,pgfplots}

\input{def}

\usetikzlibrary{shapes,arrows,positioning}
\usetikzlibrary{calc}
\usetikzlibrary{plotmarks}

\usepackage{fixmath}
\usepackage{amsmath}
\usepackage{amssymb}
\usepackage{mathrsfs}

\usepackage{algorithm}
\usepackage[noend]{algpseudocode}

\usepackage[export]{adjustbox}

\makeatletter
\def\BState{\State\hskip-\ALG@thistlm}
\makeatother

\floatname{algorithm}{Algorithm}

\usepackage{units}
\usepackage{mathtools}
\usepackage{dsfont}
\usepackage{booktabs}
\usepackage{bbm}
\usepackage{soul}
\IEEEoverridecommandlockouts

\newif\ifextendedversion 

\newif\ifmargincomments 
\margincommentstrue

\ifmargincomments

\else

\fi

\begin{document}

	\title{ \bf Energy-optimal Design and Control of Electric Powertrains\\ under Motor Thermal Constraints}


	\author{Mouleeswar Konda, Theo Hofman, Mauro Salazar%
	\thanks{The authors are with the Control Systems Technology group, Eindhoven University of Technology (TU/e), Eindhoven, 5600 MB, The Netherlands, \tt\small{k.mouleeswar@gmail.com, \{t.hofman,m.r.u.salazar\}@tue.nl}}
    }

	\maketitle
	\thispagestyle{plain}
	\pagestyle{plain}

	\begin{abstract}
	This paper presents a modeling and optimization framework to minimize the energy consumption of a fully electric powertrain by optimizing its design and control strategies whilst explicitly accounting for the thermal behavior of the Electric Motor (EM). Specifically, we first derive convex models of the powertrain components, including the battery, the EM, the transmission and a Lumped Parameter Thermal Network (LPTN) capturing the thermal dynamics of the EM. Second, we frame the optimal control problem in time domain, and devise a two-step algorithm to accelerate convergence and efficiently solve the resulting convex problem via nonlinear programming. Subsequently, we present a case study for a compact family car, optimize its transmission design and operation jointly with the regenerative braking and EM cooling control strategies for a finite number of motors and transmission technologies.
    We validate our proposed models using the high-fidelity simulation software Motor-CAD, showing that the LPTN quite accurately captures the thermal dynamics of the EM, and that the permanent magnets' temperature is the limiting factor during extended driving.
    Furthermore, our results reveal that powertrains equipped with a continuously variable transmission (CVT) result into a lower energy consumption than with a fixed-gear transmission (FGT), as a CVT can lower the EM losses, resulting in lower EM temperatures. Finally, our results emphasize the significance of considering the thermal behavior when designing an EM and the potential offered by CVTs in terms of downsizing.
    
    
	 
	\end{abstract}

	%

	\input{chapters/introduction}
	\input{chapters/methodology_1_objective}

	\input{chapters/methodology_2_vehicle_dynamics}
	\input{chapters/methodology_3_electric_motor}

	\input{chapters/methodology_4_motor_thermal}
	\input{chapters/methodology_5_battery}

	\input{chapters/methodology_6_performance_req}

	\input{chapters/methodology_7_algorithm}
	\ifextendedversion
\else
	\input{chapters/results}
	\fi
	\input{chapters/conclusion}

	\section*{Acknowledgment}
	\noindent
	We thank Dr. Ilse New for proofreading this paper. This publication is part of the NEON project with project number 17628 of the research programme Crossover which is (partly) financed by the Dutch Research Council (NWO).

	


	
	
	%
	
	\bibliographystyle{IEEEtran}        
	\bibliography{../../../Bibliography/main,../../../Bibliography/SML_papers}

\end{document}

%% file: def.tex
\newcommand{\sR}{\mathbb{R}}


%% file: chapters/introduction.tex
\section{Introduction}\label{sec:introduction}

\lettrine{T}{he} automotive industry is transitioning to electrified powertrains for several reasons, including environmental pollution and natural resource depletion. Whilst combustion engine cars are being hybridized, fully electric vehicles are slowly pervading the market. This trend is visible in all vehicle classes, from light passenger vehicles to micromobility, SUVs to long-haul trucks, electric sportscars, and motorsports~\cite{BunsenCazzolaEtAl2018,KorziliusBorsboomEtAl2021}. 
\ifextendedversion
\else
	Furthermore, electric vehicles (EVs) are becoming more practical as they provide longer ranges and superior performance compared to conventional vehicles.
\fi


However, to achieve the best market penetration of passenger electric vehicles, costs must be further reduced, which can be accomplished by downsizing the powertrain components to lower component costs and increasing the overall powertrain efficiency to lower operational costs. In this regard, the efficient downsizing of an Electric Motor (EM) necessitates for the effective use of the EM's peak performance envelope. Despite its benefits, EMs can sustain peak performance only for a limited time, due to overheating. In addition, the components are prone to failure because of the difficulties involved in cooling the downsized components. Therefore, the thermal behavior should be considered when designing an electric powertrain and its control strategies.

The thermal management system of an EV predominantly has two thermal circuits to cool its components: a low-temperature circuit for the battery and a high-temperature circuit for the inverter-motor-transmission assembly~\cite{TianWeiEtAl2018}. This research focuses on the latter. 
\ifextendedversion
\else
	Specifically, we model the thermal behavior of an EM by first capturing the temperatures of each of its subcomponents (magnets, windings, etc.) together with highly accurate loss models.
\fi
In addition, we propose a framework based on convex optimization to efficiently solve the optimization problem while controlling, first, the transmission ratio to ensure efficient operation of the EM and, second, the amount of regenerative braking to keep the EM temperatures under their limits.

\begin{figure}[!t]
	\centering
	\includegraphics[trim=0cm 0cm 0cm 0cm, clip=true,  width=\columnwidth]{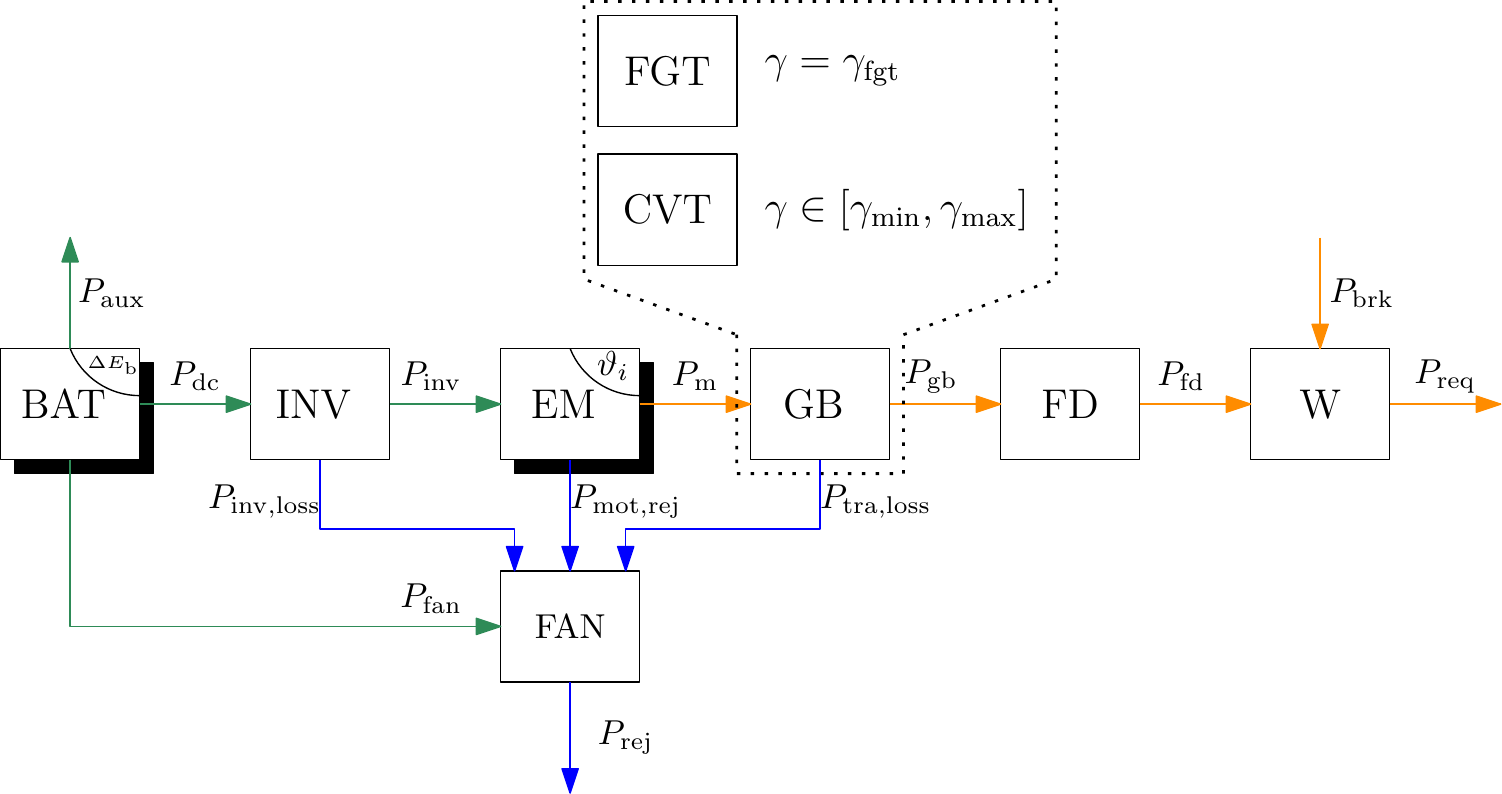}
	\caption{Schematic layout of the electric powertrain. It consists of a battery pack (BAT), an inverter (INV), an electric motor (EM), a radiator-fan assembly (FAN), a transmission (GB) which is either a fixed-gear transmission (FGT) or a continuously variable transmission (CVT); and a final drive reduction gear (FD) connected to the wheels (W). The arrows indicate the power flows between the components with mechanical power in orange, electric power in green and thermal power in blue.}
	\label{fig:Topology}
\end{figure}%

\textit{Related Literature:} The problem studied in this work pertains to two main research lines: The first one is devoted to the design and control of (hybrid) electric vehicles. The nonlinear nature of the problem was addressed through high-fidelity modeling and derivative-free methods in~\cite{EbbesenDoenitzEtAl2012,MorozovHumphriesEtAl2019,VerbruggenSilvasEtAl2020}. In~\cite{MurgovskiJohannessonEtAl2015,PourabdollahSilvasEtAl2015,BorsboomFahdzyanaEtAl2021}, convex optimization, which partly sacrifices accuracy for globally optimal and time-efficient algorithms, was used. Nonetheless, neither of the methods account for the thermal behavior of an EM at a subcomponent level. An exception is made for our previous work~\cite{LocatelloKondaEtAl2020}, which is however tailored for racing applications and hence based on less accurate models, and requires ad-hoc solution schemes that do not provide global optimality guarantees, as the underlying optimization problem is not entirely convex.

The second stream pertains to the thermal modeling of the electric machines. This problem is usually addressed with Finite-Element Analysis~(FEA), Computational Fluid Dynamics~(CFD), or Lumped Parameter Thermal Networks~(LPTNs)~\cite{BogliettiCavagninoEtAl2009}. Whilst FEA and CFD provide accurate results~\cite{KapatralIqbalEtAl2020,DongHuangEtAl2014}, they are computationally expensive and infeasible for optimization. The second method is to derive LPTNs based on first principles. These are sufficiently accurate and computationally inexpensive, which makes them suitable for optimization applications. Detailed component level LPTNs, on the other hand, have not been used in powertrain energy minimization problems. In conclusion, to the best of the authors' knowledge, there are no globally optimal methods to design and control electric powertrains accounting for their performance requirements and explicitly considering the thermal behavior of the EM.

\textit{Statement of Contributions:}
Against this backdrop, our paper presents a convex optimization framework to jointly optimize the transmission design and operation of an electric powertrain.
First, we identify a convex model of a central-EM powertrain, capturing the losses of its components, the EM temperatures and the radiator fan operation.
\ifextendedversion
\else
	Second, we validate our methods using the high-fidelity simulation software Motor-CAD~\cite{MotorCAD}. 
\fi
Finally, we perform a case study with a compact family car and compare the results for the powertrain equipped with three different motors and two transmissions: a fixed-gear transmission (FGT) and a continuously variable transmission (CVT).

\textit{Organization:} The remainder of this paper is organized as follows: Section~\ref{sec:methodology} presents the convex model of the EV powertrain, including the loss models of the EM in Section~\ref{sec:methodology_motor} and a detailed lumped parameter thermal model in Section~\ref{sec:methodology_thermal}. The optimal control problem is framed in Section~\ref{sec:methodology_optimization_problem}. The numerical results of the optimization problem are presented and analyzed in Section~\ref{sec:results}. Finally, we draw conclusions and discuss future research directions in Section~\ref{sec:conclusion}.

%% file: chapters/methodology_1_objective.tex
\section{Methodology}\label{sec:methodology}
In this section, we present a framework based on convex optimization to optimize the design and control strategies of the electric vehicle powertrain shown in Fig.~\ref{fig:Topology}. First, we define the optimization objective and frame the optimal control problem in time domain. Second, we introduce the convex quasi-static models of the powertrain components, including the transmission and battery. The temperature-dependent motor loss models and thermal models are presented in Sections~\ref{sec:methodology_motor} and~\ref{sec:methodology_thermal}, respectively. Finally, we summarize the optimization problem and discuss some key features of the proposed framework.


The all-electric powertrain consists of a battery that converts chemical energy to electrical energy. The inverter-motor assembly, in turn, converts it into mechanical energy, which is transferred from the motor shaft to the wheels through a gearbox and a final drive. We consider two gearboxes: an FGT and a CVT. The losses in the inverter, motor and transmission generate heat which is continuously removed from the powertrain to prevent any damage to the vehicle via a thermal circuit comprising the radiator, fan and the coolant. The coolant transfers the thermal energy from different components to the radiator-fan assembly, which dissipates the heat into the atmosphere.
An electric motor is capable of recuperating a part of the kinetic energy when the vehicle is decelerating which is otherwise lost to friction. The recuperation process induces losses in the motor which increases the temperatures of its components. Therefore, we regulate the temperatures by controlling the amount of power recuperated and divert the additional power to mechanical brakes.
Thereby, we consider the trade-off between maximizing regenerative braking at the cost of a higher heat generation and minimizing the radiator operation at the cost of less cooling.

The input variables to the optimal control problem with the CVT are the mechanical brake force $P_\mathrm{brk}(t)$ and the gear ratio $\gamma(t)$. For the FGT-equipped vehicle, the only input variable is $P_\mathrm{brk}(t)$, whereas the gear ratio $\gamma$ is a design variable. The state variables are the amount of battery energy used since the start of the driving cycle $\Delta E_\mathrm{b}(t)$, and the components' temperatures: the shaft's temperature $\vartheta_\mathrm{sft}(t)$, the rotor's temperature $\vartheta_\mathrm{rtr}(t)$, the permanent magnets' temperature $\vartheta_\mathrm{mgt}(t)$, the stator's temperature $\vartheta_\mathrm{str}(t)$, the windings' temperature $\vartheta_\mathrm{wdg}(t)$, and the end-windings' temperature $\vartheta_\mathrm{ewdg}(t)$. The optimal control problem will be discussed in detail below.

\subsection{Objective}\label{sec:methodology_objective}
Our objective is to minimize the internal energy consumption of the battery over a given drive cycle:
\par\nobreak\vspace{-5pt}
\begingroup
\allowdisplaybreaks 
\begin{small}
	\begin{equation}
		\label{eq:objective}
		\text{min}~J =\text{min}~\Delta E_{\mathrm{b}},
	\end{equation}
\end{small}%
\endgroup
where $\Delta E_{\mathrm{b}}$ is the difference in the battery state of energy defined as
\par\nobreak\vspace{-5pt}
\begingroup
\allowdisplaybreaks 
\begin{small}
	\begin{equation}
		\label{eq:DeltaEb}
		\Delta E_{\mathrm{b}} = E_{\mathrm{b}}(0) - E_{\mathrm{b}}(\mathrm{T}),
	\end{equation}
\end{small}%
\endgroup
and $E_{\mathrm{b}}(0)$ and $E_{\mathrm{b}}(\mathrm{T})$ denote the battery state of energy at the start and end of the drive cycle, respectively.


%% file: chapters/methodology_2_vehicle_dynamics.tex
\subsection{Vehicle Dynamics and Transmission}\label{sec:methodology_vehicle}
In this section, we model the vehicle and the transmission in a quasi-static manner in line with current practices~\cite{GuzzellaSciarretta2007}. In this regard, we present a convex model of the longitudinal vehicle dynamics in time domain. To improve readability, we exclude time dependence whenever it is clear from the context. First, the power equilibrium at the wheels is given as
\par\nobreak\vspace{-5pt}
\begingroup
\allowdisplaybreaks 
\begin{small}
	\begin{equation}
		\label{eq:Power_Balance_FD}
		P_{\mathrm{fd}} = P_{\mathrm{req}} + P_{\mathrm{brk}},
	\end{equation}
\end{small}%
\endgroup
\noindent where $P_{\mathrm{fd}}$ is the final drive power, $P_{\mathrm{req}}$ is the power required at the wheels and $P_{\mathrm{brk}}$ is the mechanical brake power required. As mentioned above, $P_{\mathrm{brk}}$ is an input variable, and it controls the amount of regenerative braking to maintain EM temperatures within their limits. Additionally, we constrain the brake power according to
\par\nobreak\vspace{-5pt}
\begingroup
\allowdisplaybreaks 
\begin{small}
	\begin{equation}
		\label{eq:Pbrk>0}
		P_{\mathrm{brk}} \geq 0.
	\end{equation}
\end{small}%
\endgroup
\noindent $P_{\mathrm{req}}$ is a combination of aerodynamic drag, rolling resistance, gravitational force and vehicle inertia. We compute $P_{\mathrm{req}}$ for a given drive cycle with a velocity $v(t)$, an acceleration $a(t)$, and a road gradient $\alpha(t)$ with
\par\nobreak\vspace{-5pt}
\begingroup
\allowdisplaybreaks 
\begin{small}
	\begin{equation}
		\label{eq:Power_Required}
		\begin{aligned}
			P_{\mathrm{req}}\big(v(t),\alpha(t),&a(t)\big) = v(t) \cdot \Big( \frac{1}{2} \cdot \rho_{\mathrm{a}} \cdot c_{\mathrm{d}} \cdot A_{\mathrm{f}} \cdot v(t)^2 + \\  &m_\mathrm{v}  \big(g \cdot c_{\mathrm{r}} \cdot \cos(\alpha(t)) + g \cdot \sin(\alpha(t)) + a(t) \big) \Big),
		\end{aligned}
	\end{equation}
\end{small}%
\endgroup
where $m_\mathrm{v}$ is the total mass of the vehicle, $c_{\mathrm{d}}$ is the vehicle's drag coefficient, $A_{\mathrm{f}}$ is the frontal area of the vehicle, $c_{\mathrm{r}}$ is the road friction coefficient, $\rho_{\mathrm{a}}$ is the air density and $g$ is the Earth's gravitational constant. We assume a constant final drive and transmission efficiency, $\eta_\mathrm{fd}$ and $\eta_\mathrm{gb}$, respectively. Therefore, the power at the motor shaft, $P_\mathrm{m}$, is given by
\par\nobreak\vspace{-5pt}
\begingroup
\allowdisplaybreaks 
\begin{small}
	\begin{equation}
		\label{eq:FD-MOT_Power}
		P_\mathrm{m} = 
		\begin{cases}
			\dfrac{1}{\eta_\mathrm{gb} \cdot \eta_\mathrm{fd}} \cdot P_\mathrm{fd} & \text{~~if~~} P_\mathrm{fd} \geq 0 \\
			\eta_\mathrm{gb} \cdot \eta_\mathrm{fd} \cdot r_\mathrm{b} \cdot P_\mathrm{fd} & \text{~~if~~} P_\mathrm{fd} < 0, \\
		\end{cases}
	\end{equation}
\end{small}%
\endgroup
where $\eta_\mathrm{gb} = \eta_\mathrm{fgt}$ and $\eta_\mathrm{gb} = \eta_\mathrm{cvt}$ are efficiencies of FGT and CVT, respectively, and $r_\mathrm{b}$ is the regenerative braking fraction. 
The rotational speed of the motor shaft for a given gear ratio $\gamma$ is calculated using
\par\nobreak\vspace{-5pt}
\begingroup
\allowdisplaybreaks 
\begin{small}
	\begin{equation}
		\label{eq:W_to_MOt_Speed}
		\omega_{\mathrm{m}} = \gamma \cdot \gamma_{\mathrm{fd}} \cdot \dfrac{v(t)}{r_{\mathrm{w}}}, 
	\end{equation}
\end{small}%
\endgroup
where $r_{\mathrm{w}}$ is the radius of the wheel, $\gamma_{\mathrm{fd}}$ is the final drive gear ratio, and $\gamma$ is the transmission gear ratio. The transmission ratio is one of the optimization variables and depending on the type of transmission, it is constrained as
\par\nobreak\vspace{-5pt}
\begingroup
\allowdisplaybreaks
\begin{small}
	\begin{equation}%
		\label{eq:gamma_Limits}
		\gamma(t) 
		\begin{cases}
			= \gamma_\mathrm{fgt} > 0&\forall~t~~\text{ if  FGT},   \\
			\in [\gamma_{\mathrm{min}}, \gamma_{\mathrm{max}}]\subset \sR_{++}  & \forall~t~~\text{ if  CVT},
		\end{cases}
	\end{equation}
\end{small}%
\endgroup
where $\sR_{++}$ is the set of positive real numbers.
Finally, the total mass of the vehicle is
\par\nobreak\vspace{-5pt}
\begingroup
\allowdisplaybreaks
\begin{small}
	\begin{equation}%
		\label{eq:Mass}
		m_{\mathrm{v}} = m_{\mathrm{0}} + m_{\mathrm{m}} +  
		\begin{cases}
			m_{\mathrm{fgt}} &\text{ if  FGT},   \\
			m_{\mathrm{cvt}} &\text{ if  CVT},  
		\end{cases}
	\end{equation}
\end{small}%
\endgroup
where $m_{\mathrm{0}}$ is the base mass of the vehicle, $m_{\mathrm{m}}$ is the mass of the motor, $m_{\mathrm{fgt}}$ is the mass of the FGT, and $m_{\mathrm{cvt}}$ is the mass of the CVT. We assume that the base mass includes the frame's mass, the battery's mass, and the equivalent mass due to the moment of inertia of rotating parts.

%% file: chapters/methodology_3_electric_motor.tex
\subsection{Electric Motor and Inverter}\label{sec:methodology_motor}
\ifextendedversion
\else
In this section, we derive an accurate temperature-dependent model of the EM losses inspired by our previous work~\cite{LocatelloKondaEtAl2020,HurkSalazar2021}. 
\fi
One of the most common types of motors for a light passenger vehicle is an Interior Permanent Magnet (IPM) motor~\cite{HwangHanEtAl2018}. We model the IPM motor based on the templates provided in the high-fidelity simulation software Motor-CAD. We use the data generated by the software to identify and validate the motor loss models and motor thermal models. Subsequently, the power at the motor terminals, $P_\mathrm{inv}$, is given by
\par\nobreak\vspace{-5pt}
\begingroup
\allowdisplaybreaks
\begin{small}
	\begin{equation}%
		\label{eq:MOT_Power_Balance}
		P_{\mathrm{inv}} = P_{\mathrm{m}} + P_{\mathrm{loss}},
	\end{equation}
\end{small}%
\endgroup
where $P_{\mathrm{loss}}$ represents the combined losses of all the motor subcomponents and can be defined as
\par\nobreak\vspace{-5pt}
\begingroup
\allowdisplaybreaks
\begin{small}
	\begin{equation}%
		\label{eq:MOT_Ploss_total}
		P_{\mathrm{loss}} = \sum_i P_i,~~~~~\forall~~~\{i = \text{sft, rtr, mgt, str, wdg}\},
	\end{equation}
\end{small}%
\endgroup
where $P_{i}$ represents the losses corresponding to the subcomponents of the motor, namely the shaft (sft), the rotor (rtr), the permanent magnets (mgt), the stator (str) and the windings (wdg). The losses are individually identified for each EM component in order to build an LPTN and explicitly capture the thermal dynamics of the EM, which will be discussed in Section~\ref{sec:methodology_thermal}. The shaft losses, $P_\mathrm{sft}$, represent the bearing friction losses and are independent of motor power and temperature. Therefore, we compute them using the linear relation
\par\nobreak\vspace{-5pt}
\begingroup
\allowdisplaybreaks 
\begin{small}
	\begin{equation}
		\label{eq:Shaft_Losses}
		\begin{aligned}
			P_\mathrm{sft} &= a_\mathrm{sft,0} + a_\mathrm{sft,1} \cdot \omega_{\mathrm{m}},
		\end{aligned}
	\end{equation}
\end{small}%
\endgroup
where $a_{\mathrm{sft}}$ is subject to identification. 
\ifextendedversion
\else
	Now we present the loss models for the other EM subcomponents for which we divide the driving cycle into two parts: traction (including coasting) and braking.
\fi
During traction, the motor must supply the full power required to propel the vehicle. In contrast, when braking, we can split the power between the friction brakes and EM (via regenerative braking) to keep the EM's temperatures within their limits. The main advantage of this model is that for a given drive cycle, the motor power is known during traction, which enables us to accurately model the EM losses by using motor-power-level-specific fitting coefficients.
To this end, we compute the minimum tractive power required to propel the vehicle, $\overline{P}_\mathrm{m}$ as
\par\nobreak\vspace{-5pt}
\begingroup
\allowdisplaybreaks 
\begin{small}
	\begin{equation}
		\label{eq:Pm_bar}
		\overline{P}_\mathrm{m} = \text{max} \Big( \dfrac{1}{\eta_\mathrm{gb} \cdot \eta_\mathrm{fd}} \cdot P_\mathrm{req}, 
		\eta_\mathrm{gb} \cdot \eta_\mathrm{fd} \cdot r_\mathrm{b} \cdot P_\mathrm{req}, {P}_\mathrm{m,min} \Big),
	\end{equation}
\end{small}%
\endgroup
where ${P}_\mathrm{m,min}$ is the minimum EM power.
First, the rotor and stator losses represent the iron losses in the EM's rotor and stator, respectively, and are computed with the relaxed equations
\par\nobreak\vspace{-5pt}
\begingroup
\allowdisplaybreaks
\begin{small}
	\begin{equation}
		\label{eq:Rtr_Loss_Complex}
		P_\mathrm{rtr} \geq \\
		\begin{cases}
			a_\mathrm{rtr,0}(\overline{P}_\mathrm{m}) + 
			a_\mathrm{rtr,1}(\overline{P}_\mathrm{m}) \cdot \omega_{\mathrm{m}} + 
			a_\mathrm{rtr,2}(\overline{P}_\mathrm{m}) \cdot \omega_{\mathrm{m}}^2,\\
			\qquad \qquad \qquad \qquad \qquad \qquad \qquad \qquad \qquad \text{if} ~ \overline{P}_\mathrm{m} \geq 0,\\
			y^\top Q_\mathrm{rtr} y, \qquad \qquad \qquad \qquad \qquad \qquad ~~~~~ \text{if} ~ \overline{P}_\mathrm{m} < 0,\\
		\end{cases}\\
	\end{equation}
\end{small}%
\endgroup
\par\nobreak\vspace{-15pt}
\begingroup
\allowdisplaybreaks
\begin{small}
	\begin{equation}
		\label{eq:Str_Loss_Complex}
		P_\mathrm{str} \geq \\
		\begin{cases} 
			a_\mathrm{str,0}(\overline{P}_\mathrm{m}) + 
			a_\mathrm{str,1}(\overline{P}_\mathrm{m}) \cdot \omega_{\mathrm{m}} + 
			a_\mathrm{str,2}(\overline{P}_\mathrm{m}) \cdot \omega_{\mathrm{m}}^2,\\
			\qquad \qquad \qquad \qquad \qquad \qquad \qquad \qquad \qquad \text{if} ~ \overline{P}_\mathrm{m} \geq 0,\\
			y^\top Q_\mathrm{str} y, \qquad \qquad \qquad \qquad \qquad \qquad ~~~~~ \text{if} ~ \overline{P}_\mathrm{m} < 0,\\
		\end{cases}\\
	\end{equation}
\end{small}%
\endgroup
\ifextendedversion
\else
	where $a_\mathrm{rtr}$, $a_\mathrm{str}$, $Q_\mathrm{rtr}$ and $Q_\mathrm{str}$ are subject to identification, and $y = \begin{bmatrix} 1 & \omega_{\mathrm{m}} & P_{\mathrm{m}} & \omega_{\mathrm{m}}^2 & P_{\mathrm{m}}^2 & \omega_{\mathrm{m}} \cdot P_{\mathrm{m}} \end{bmatrix} ^\top$. To retain convexity in the positive speed domain, we ensure that $a_\mathrm{rtr,2}~\geq~0$, $a_\mathrm{str,2}~\geq~0$, and $Q_\mathrm{rtr}$, $Q_\mathrm{str}$ are positive semi-definite matrices. Given the objective~\eqref{eq:objective}, constraints~\eqref{eq:Rtr_Loss_Complex} and~\eqref{eq:Str_Loss_Complex} will always hold with equality at the optimum~\cite{VerbruggenSalazarEtAl2019}. For the remainder of this paper we will directly introduce all constraints in their convex relaxed form, following the same rationale.
\fi
Second, the magnet loss models are temperature-dependent and computed using convex quadratic relations of the form
\par\nobreak\vspace{-5pt}
\begingroup
\allowdisplaybreaks
\begin{small}
	\begin{equation}
		\label{eq:Mgt_Loss_Complex}
		P_\mathrm{mgt} \geq 
		\begin{cases}
			z_\mathrm{mgt}^\top~R_\mathrm{mgt}(\overline{P}_\mathrm{m})~z_\mathrm{mgt}, &\quad \overline{P}_\mathrm{m} \geq 0,\\
			y^\top Q_\mathrm{mgt} y, &\quad \overline{P}_\mathrm{m} < 0,
		\end{cases}
	\end{equation}
\end{small}%
\endgroup
\ifextendedversion
\else
	where $Q_\mathrm{mgt}$ and $R_\mathrm{mgt}$ are positive semi-definite matrices subject to identification, and $z_\mathrm{mgt} = \begin{bmatrix} 1 & \omega_{\mathrm{m}} & \vartheta_{\mathrm{mgt}} \end{bmatrix} ^\top$.
\fi
Third, the copper losses in the EM are represented by temperature-dependent winding losses as
\par\nobreak\vspace{-5pt}
\begingroup
\allowdisplaybreaks
\begin{small}
	\begin{equation}
		\label{eq:Wdg_Loss_Complex}
		P_\mathrm{wdg} \geq \\ 
		\begin{cases}
			z_\mathrm{wdg}^\top~R_\mathrm{wdg}(\overline{P}_\mathrm{m})~z_\mathrm{wdg} + 
			\dfrac{a_\mathrm{wdg,1}(\overline{P}_\mathrm{m})}{\omega_\mathrm{m}} + 
			\dfrac{a_\mathrm{wdg,2}(\overline{P}_\mathrm{m})}{\omega_\mathrm{m}^2}, \\
			\qquad \qquad \qquad \qquad \qquad \qquad \qquad \qquad \qquad \text{if} ~ \overline{P}_\mathrm{m} \geq 0,\\
			y^\top Q_\mathrm{wdg} y, \qquad \qquad \qquad \qquad \qquad \qquad ~~~~ \text{if} ~ \overline{P}_\mathrm{m} < 0,\\
		\end{cases}	\\	
	\end{equation}
\end{small}%
\endgroup
\ifextendedversion
\else
	where $a_\mathrm{wdg}$, $Q_\mathrm{wdg}$, and $R_\mathrm{wdg}$ are subject to identification and $z_\mathrm{wdg} = \begin{bmatrix} 1 & \omega_{\mathrm{m}} & \vartheta_{\mathrm{wdg}} & \omega_{\mathrm{m}}^2 & \vartheta_{\mathrm{wdg}}^2 & \omega_{\mathrm{m}} \cdot \vartheta_{\mathrm{wdg}} \end{bmatrix} ^\top$. In addition, we ensure that   $a_\mathrm{wdg,1} \geq 0$, $a_\mathrm{wdg,2} \geq 0$, and $Q_\mathrm{wdg}$, $R_\mathrm{wdg}$ are positive semi-definite matrices to preserve convexity in the positive speed domain.
\fi
To prevent the winding losses from reaching infinity when the vehicle is starting, we set $a_\mathrm{wdg,1} = 0$ and $a_\mathrm{wdg,2} = 0$ at very low speeds. 
\ifextendedversion
\else
Moreover, we constrain the power losses with 
\fi
\par\nobreak\vspace{-5pt}
\begingroup
\allowdisplaybreaks 
\begin{small}
	\begin{equation}
		\label{eq:MOT_Loss_geq0}
		P_{i} \geq 0, ~~~~~\forall~~~\{i = \text{sft, rtr, mgt, str, wdg}\}.
	\end{equation}
\end{small}%
\endgroup
\begin{figure}[t!]	
	\centering
	\includegraphics[trim=1.17cm 0.4cm 1.65cm 0.45cm, clip=true, width=\columnwidth]{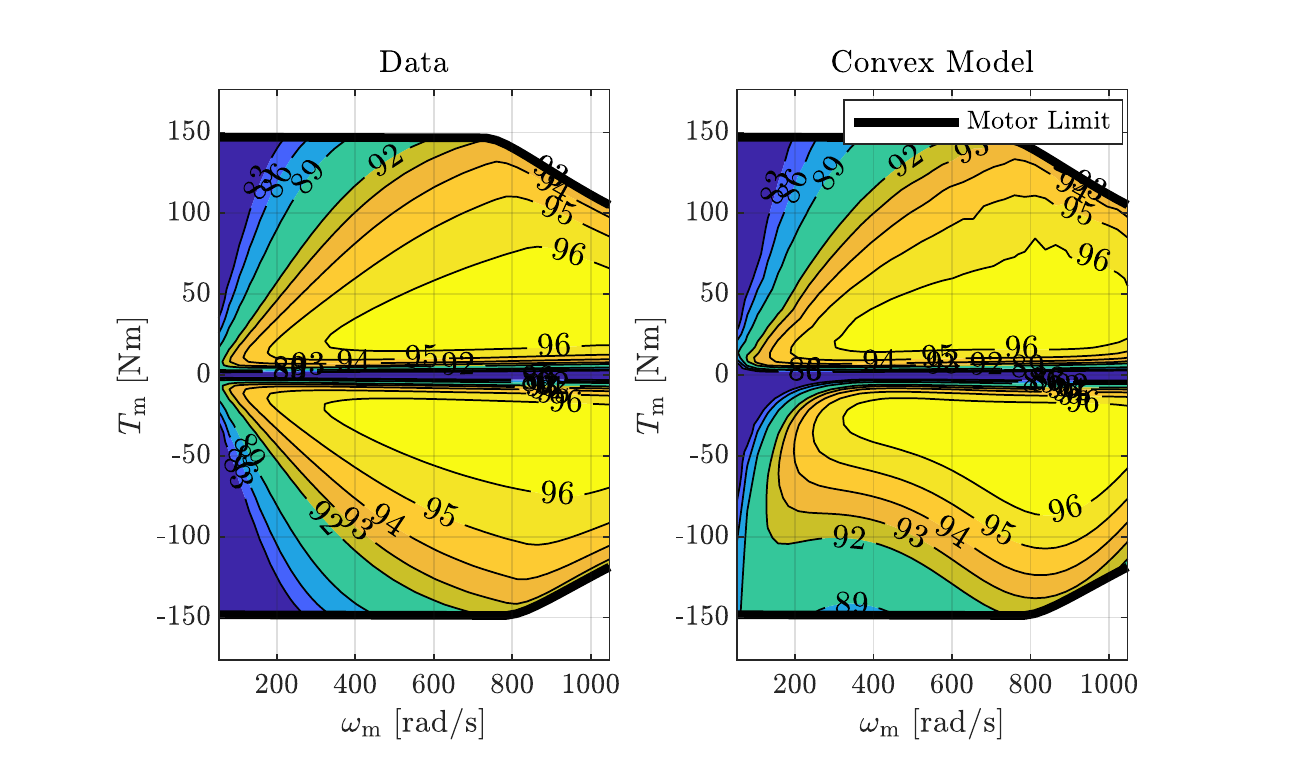}
	\caption{Efficiency map of the interior permanent magnet (IPM) machine (left) and its convex model (right) at a temperature of \unit[65]{$^{\circ}$C}. }
	\label{fig:Efficiency_Data_Model}
\end{figure}%
\ifextendedversion
\else
Fig.~\ref{fig:Efficiency_Data_Model} compares the efficiency map obtained from Motor-CAD with the efficiency map from convex models. We can observe that the models are accurate, especially in the positive domain where most of the operation occurs.
\fi
The torque limits of the EM are 
\par\nobreak\vspace{-5pt}
\begingroup
\allowdisplaybreaks 
\begin{small}
	\begin{equation}
		\label{eq:MOT_Torque_Limit}
		P_{\mathrm{m}} \in \left[T_\mathrm{m,min}, T_\mathrm{m,max}\right] \cdot \omega_{\mathrm{m}},
	\end{equation}
\end{small}%
\endgroup
\ifextendedversion
\else
where $T_\mathrm{m,max}$ and $T_\mathrm{m,min}$ are the maximum and minimum torques subject to identification.
\fi
In addition, we bound the EM power as 
\par\nobreak\vspace{-5pt}
\begingroup
\allowdisplaybreaks 
\begin{small}
	\begin{equation}
		\label{eq:MOT_Power_Limit}
		\begin{aligned}
			P_{\mathrm{m}} &\in \left[P_{\mathrm{m,min}}, P_{\mathrm{m,max}}\right],\\
		\end{aligned}		
	\end{equation}
\end{small}%
\endgroup
\ifextendedversion
\else
where $P_{\mathrm{m,min}}$ and $P_{\mathrm{m,max}}$ are the minimum and maximum motor powers, respectively. We compute them using the relationships $P_{\mathrm{m,min}}~=~d_\mathrm{1,min}~\cdot~\omega_{\mathrm{m}}~+~d_\mathrm{0,min}$, $P_{\mathrm{m,max}}~=~d_\mathrm{1,max}~\cdot~\omega_{\mathrm{m}}~+~d_\mathrm{0,max}$ where $d_\mathrm{0,min}$, $d_\mathrm{1,min}$, $d_\mathrm{0,max}$ and $d_\mathrm{1,max}$ are the coefficients subject to identification.
\fi
The non-negative rotational speed of the motor is bounded by the maximum EM speed, $\omega_{\mathrm{m,max}}$, as
\par\nobreak\vspace{-5pt}
\begingroup
\allowdisplaybreaks 
\begin{small}
	\begin{equation}
		\label{eq:MOT_Speed_Limit}
		\omega_{\mathrm{m}} \in \left[0, \omega_{\mathrm{m,max}}\right].
	\end{equation}
\end{small}%
\endgroup
Finally, we approximate and relax the inverter losses using the quadratic function
\par\nobreak\vspace{-5pt}
\begingroup
\allowdisplaybreaks 
\begin{small}
	\begin{equation}
		\label{eq:INV_BAT_Losses}
		P_{\mathrm{dc}} \geq	\alpha_\mathrm{inv} \cdot P_{\mathrm{inv}}^2 + P_{\mathrm{inv}},
	\end{equation}
\end{small}%
\endgroup
where $\alpha_\mathrm{inv}\geq0$ is the inverter loss-coefficient subject to identification,
\ifextendedversion
\else
and $P_{\mathrm{dc}}$ is the power at the inverter terminals.
\fi

%% file: chapters/methodology_4_motor_thermal.tex
\subsection{Motor Thermal Model and Fan Model}\label{sec:methodology_thermal}
\begin{figure}[t!]
	\centering
	\includegraphics[trim=0.5cm 0.25cm 0.25cm 0cm,clip=true, width=\columnwidth]{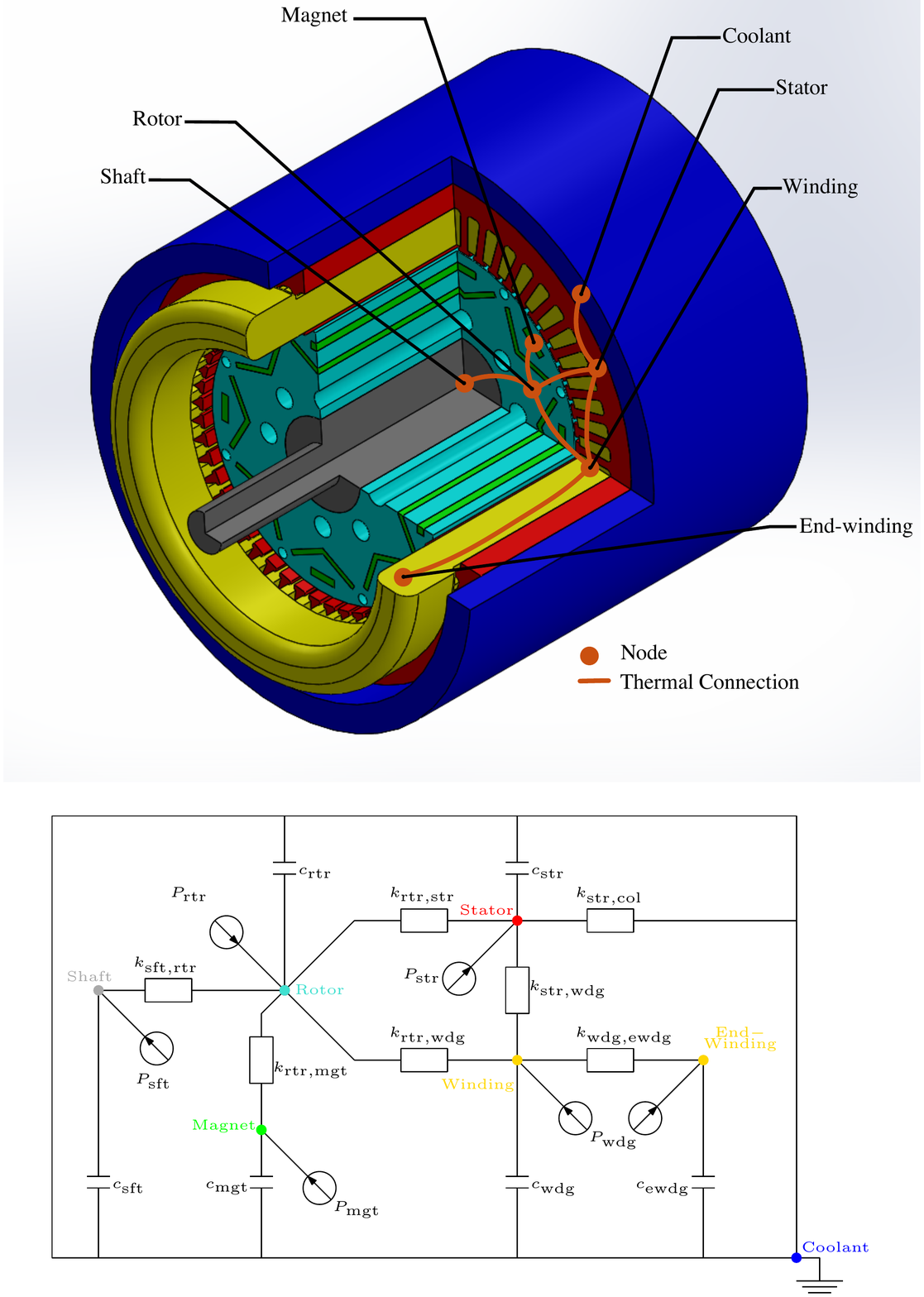}
	\caption{A 3-dimensional section view of the IPM motor from Motor-CAD (top) and the corresponding Lumped Parameter Thermal Network (bottom).}
	\label{fig:ThermalNetwork}
\end{figure}%
In this section, we derive an LPTN model, based on~\cite{LocatelloKondaEtAl2020,WeiHofmanEtAl2021,WangJagarwalEtAl2015}. In addition to the EM components presented in Section~\ref{sec:methodology_motor}, we include the overhanging copper windings, further referred to as end-windings (edwg), in our thermal model because the end-windings reach higher temperatures and are the limiting factors in high-performance operations~\cite{MadonnaWalkerEtAl2018}. Therefore, we build the LPTN with a total of 6 nodes: the shaft (sft), the rotor (rtr), the permanent magnets (mgt), the stator (str), the windings (wdg) and the end-windings (edwg). The LPTN is based on the following assumptions: First, the heat flow in the radial direction is independent of the heat flow in the axial direction. Second, the heat flow in the circumferential direction is absent. Third, each component's thermal properties, including its temperature, can be represented by a single node, i.e., the temperature distribution in a component is uniform. Fig.~\ref{fig:ThermalNetwork} shows the cross-section of the IPM machine at the top and its LPTN at the bottom. The energy balance equations for the LPTN are given by 
\par\nobreak\vspace{-5pt}
\begingroup
\allowdisplaybreaks 
 \begin{small}
	\begin{equation}
		\begin{aligned}
			P_{\mathrm{sft}} &= c _{\mathrm{sft}}  \dot{\vartheta}_{\mathrm{sft}} + 
								  k_{\mathrm{sft,rtr}} (\vartheta_{\mathrm{sft}} - \vartheta_{\mathrm{rtr}}) \\
			P_{\mathrm{rtr}} &= c _{\mathrm{rtr}}  \dot{\vartheta}_{\mathrm{rtr}} + 
								  k_{\mathrm{sft,rtr}} (\vartheta_{\mathrm{rtr}} - \vartheta_{\mathrm{sft}}) + 
								  k_{\mathrm{rtr,mgt}} (\vartheta_{\mathrm{rtr}} - \vartheta_{\mathrm{mgt}})  \\  
							   &  \quad + k_{\mathrm{rtr,str}} (\vartheta_{\mathrm{rtr}} - \vartheta_{\mathrm{str}}) + 
								  k_{\mathrm{rtr,wdg}} (\vartheta_{\mathrm{rtr}} - \vartheta_{\mathrm{wdg}})\\
			P_{\mathrm{mgt}} &= c_{\mathrm{mgt}}  \dot{\vartheta}_{\mathrm{mgt}} + 
							      k_{\mathrm{rtr,mgt}} (\vartheta_{\mathrm{mgt}} - \vartheta_{\mathrm{rtr}}) \\
			P_{\mathrm{str}} &= c_{\mathrm{str}}  \dot{\vartheta}_{\mathrm{str}} + 
								  k_{\mathrm{rtr,str}} (\vartheta_{\mathrm{str}} - \vartheta_{\mathrm{rtr}}) + 
								  k_{\mathrm{str,wdg}} (\vartheta_{\mathrm{str}} - \vartheta_{\mathrm{wdg}}) \\ 
							   &  \quad + k_{\mathrm{str,col}} (\vartheta_{\mathrm{str}} - \vartheta_{\mathrm{col}})\\
			P_{\mathrm{wdg}} &= c_{\mathrm{wdg}}  \dot{\vartheta}_{\mathrm{wdg}} + 
								  k_{\mathrm{str,wdg}} (\vartheta_{\mathrm{wdg}} - \vartheta_{\mathrm{str}})\\
							   &  \quad + k_{\mathrm{rtr,wdg}} (\vartheta_{\mathrm{wdg}} - \vartheta_{\mathrm{rtr}}) + 
								  k_{\mathrm{wdg,ewdg}} (\vartheta_{\mathrm{wdg}} - \vartheta_{\mathrm{ewdg}})\\
			P_{\mathrm{wdg}} &= c_{\mathrm{ewdg}}  \dot{\vartheta}_{\mathrm{ewdg}} + 
								  k_{\mathrm{wdg,ewdg}} (\vartheta_{\mathrm{ewdg}} - \vartheta_{\mathrm{wdg}}),
		\end{aligned}	
	\end{equation}
\end{small}%
\endgroup
where $\vartheta_i$ and $\dot{\vartheta}_i$ represent the temperature of a node and the rate of change of temperature, respectively, for the nodes, $i \in \{ \text{sft, rtr, mgt, str, wdg, ewdg} \} $. Parameter $k_{i,j}$ represents the overall heat transfer coefficient between node $i$ and node $j$, and $c_i$ represents the heat capacity of the node.  Parameters $k_{i,j}$ and $c_i$ are subject to identification. We use nonlinear gradient-based methods to identify the thermal coefficients. In addition, we use the losses estimated by our power loss models instead of the losses from the high-fidelity software to avoid propagating the error of the power loss model to the LPTN.
In order to prevent motor failure due to thermal limitations, we constrain the temperature of each node using
\par\nobreak\vspace{-5pt}
\begingroup
\allowdisplaybreaks 
\begin{small}
	\begin{equation}
		\label{eq:Thermal_Limits}
		\vartheta_i \leq \vartheta_{i,\mathrm{max}},
	\end{equation}
\end{small}%
\endgroup
where $\vartheta_{i,\mathrm{max}}$ is the temperature limit of node $i$. We initialize the temperature of each node at the coolant temperature, $\vartheta_{\mathrm{col}}$, as
\par\nobreak\vspace{-5pt}
\begingroup
\allowdisplaybreaks 
\begin{small}
	\begin{equation}
		\label{eq:T_start}
		\vartheta_{i}(0) = \vartheta_{\mathrm{col}}.
	\end{equation}
\end{small}%
\endgroup
Finally, the air flow rate, $\dot{m}_{\mathrm{air}}$, required by the fan to remove heat is given by
\par\nobreak\vspace{-5pt}
\begingroup
\allowdisplaybreaks 
\begin{small}
	\begin{equation}
		\label{eq:FAN_air}
		\dot{m}_{\mathrm{air}} = \frac{ P_\mathrm{mot,rej} + 	P_\mathrm{loss,inv} + P_\mathrm{loss,tra} }{\eta_\mathrm{he} \cdot C_\mathrm{p,air} \cdot \Delta \vartheta_{\mathrm{a}}},
	\end{equation}
\end{small}%
\endgroup
where the transmission losses are $P_\mathrm{loss,tra}~=~P_\mathrm{m}~-~P_\mathrm{fd}$, the inverter losses are given as $P_\mathrm{loss,inv}~=~P_\mathrm{dc}~-~P_\mathrm{inv}$ and the heat rejected by the EM is $P_\mathrm{mot,rej}~=~k_{\mathrm{str,col}}~(\vartheta_{\mathrm{str}}~-~\vartheta_{\mathrm{col}})$. In addition, $\eta_\mathrm{he}$ is the heat exchanger efficiency, $C_\mathrm{p,air}$ is the specific heat capacity of air and $\Delta \vartheta_{\mathrm{a}}$ is the constant temperature gain of the air across the radiator~\cite{WangJagarwalEtAl2015}. We compute the power required by the fan, $P_\mathrm{fan}$ as
\par\nobreak\vspace{-5pt}
\begingroup
\allowdisplaybreaks 
\begin{small}
	\begin{equation}
		\label{eq:FAN_Power}
		P_{\mathrm{fan}} \geq \alpha_{\mathrm{f}} \cdot \dot{m}_{\mathrm{air}}^2,
	\end{equation}
\end{small}%
\endgroup
\ifextendedversion
\else
where $\alpha_{\mathrm{f}}\geq 0$ is again subject to identification. Finally, the air flow rate is constrained as
\fi
\par\nobreak\vspace{-5pt}
\begingroup
\allowdisplaybreaks 
\begin{small}
	\begin{equation}
		\label{eq:Max_Fan_Flow}
		\dot{m}_{\mathrm{air}} \leq \dot{m}_{\mathrm{air,max}},
	\end{equation}
\end{small}%
\endgroup
where the maximum air flow rate, $\dot{m}_{\mathrm{air,max}}$, is a given parameter.
Hereby, we observe that $P_\mathrm{fan}$ is not an explicit control variable, but rather results from the airflow needed to guarantee a constant temperature gain $\Delta \vartheta_{\mathrm{a}}$ for the given losses, as expressed in~\eqref{eq:FAN_air}.

%% file: chapters/methodology_5_battery.tex
\subsection{Battery}\label{sec:methodology_battery}
\ifextendedversion
\else
	In this section, we derive a detailed energy-dependent battery model in line with~\cite{VerbruggenSalazarEtAl2019}.
\fi
First, the electric power at the battery terminals, $P_{\mathrm{bat}}$, is computed as
\par\nobreak\vspace{-5pt}
\begingroup
\allowdisplaybreaks 
\begin{small}
	\begin{equation}
		\label{eq:INV_BAT_AUX_Equillibrium}	
		P_{\mathrm{bat}} = P_{\mathrm{dc}} + P_{\mathrm{aux}} + P_{\mathrm{fan}},
	\end{equation}
\end{small}%
\endgroup

\ifextendedversion
\else
	\noindent where $P_{\mathrm{aux}}$ is the auxiliary power. Subsequently, the internal battery power $P_{\mathrm{int}}$, which changes the battery state of energy, is related to $P_{\mathrm{bat}}$ as
\fi

\par\nobreak\vspace{-5pt}
\begingroup
\allowdisplaybreaks 
\begin{small}
	\begin{equation}
		\label{eq:BAT_INT_Losses_Complex}
		(P_{\mathrm{int}} - P_{\mathrm{bat}}) \cdot P_{\mathrm{oc}}\geq P_{\mathrm{int}}^2,
	\end{equation}
\end{small}%
\endgroup
where $P_{\mathrm{oc}}$ is the open circuit power dependent on the internal resistance of the battery and its open circuit voltage. Furthermore, $P_{\mathrm{oc}}$ is a function of the state of energy (SoE) of the battery and defined as
\par\nobreak\vspace{-5pt}
\begingroup
\allowdisplaybreaks 
\begin{small}
	\begin{equation}
		\label{eq:Poc_Linear_Eq}
		P_{\mathrm{oc}} = b_\mathrm{1} \cdot E_{\mathrm{b}} + b_\mathrm{2} \cdot E_{\mathrm{b,max}},
	\end{equation}
\end{small}%
\endgroup
where $b_\mathrm{1}$ and $b_\mathrm{2}$ are subject to identification~\cite{VerbruggenSalazarEtAl2019}. 
Additionally, we bound the battery SoE using the minimum and maximum state of charge (SoC) levels, $\zeta_{\mathrm{b,min}}$ and $\zeta_{\mathrm{b,max}}$, respectively, as
\par\nobreak\vspace{-5pt}
\begingroup
\allowdisplaybreaks 
\begin{small}
	\begin{equation}
		\label{eq:BAT_SOC_Limit}
		E_\mathrm{b} \in \left[ \zeta_{\mathrm{b,min}}, \zeta_{\mathrm{b,max}}\right] \cdot E_{\mathrm{b,max}}.
	\end{equation}
\end{small}%
\endgroup
We assume that the vehicle starts with a full battery at the start of the cycle:
\par\nobreak\vspace{-5pt}
\begingroup
\allowdisplaybreaks 
\begin{small}
	\begin{equation}
		\label{eq:Battery_Initial}
		E_\mathrm{b}(0) = E_\mathrm{b,max} \cdot \zeta_{\mathrm{b,max}}.
	\end{equation}
\end{small}%
\endgroup
Finally, the battery SoE changes with $P_\mathrm{int}$ as
\par\nobreak\vspace{-5pt}
\begingroup
\allowdisplaybreaks 
\begin{small}
	\begin{equation}
		\label{eq:Battery_Dynamics}
		\frac{\mathrm{d}}{\mathrm{d}t}E_\mathrm{b} = -P_\mathrm{int}.
	\end{equation}
\end{small}%
\endgroup

%% file: chapters/methodology_6_performance_req.tex
\subsection{Performance Requirements}\label{sec:methodology_performance}
In this section, we derive the performance requirements of the vehicle in order to ascertain that they are within acceptable limits. In line with~\cite{VerbruggenSalazarEtAl2019}, we capture the gradeability requirement as
\par\nobreak\vspace{-5pt}
\begingroup
\allowdisplaybreaks
\begin{small}
	\begin{equation}
		m_\mathrm{v}  \cdot g \cdot \sin{(\alpha_\mathrm{start})} \cdot r_\mathrm{w} \leq T_\mathrm{m,max} \cdot \eta_\mathrm{fd} \cdot \gamma_{\mathrm{fd}} \left\{\begin{array}{ll}
			\eta_\mathrm{fgt} \cdot \gamma_1 & \text{ if FGT }, \\ 
			\eta_\mathrm{cvt} \cdot \gamma_\mathrm{max} & \text{ if CVT},
		\end{array}\right.
		\label{eq:SlopeStartConstraint}
	\end{equation}
\end{small}%
\endgroup
where $\alpha_\mathrm{start}$ is the required starting gradient. 
Lastly, in line with~\cite{KorziliusBorsboomEtAl2021}, we ensure that the EM can deliver the required torque to propel the vehicle at its top speed on a flat road using the constraint
\par\nobreak\vspace{-5pt}
\begingroup
\allowdisplaybreaks
\begin{small}
	\begin{equation}
		\label{eq:TopSpeedConstraint}
			\begin{aligned}[b]
			T_\mathrm{m,speed} \leq \min(&T_\mathrm{em,max}\cdot \eta_\mathrm{fd}\cdot \eta_\mathrm{gb}\cdot \gamma_\mathrm{x}\cdot \gamma_\mathrm{fd}, \\
			& (d_\mathrm{1,max}\cdot\gamma_\mathrm{x}\cdot \gamma_\mathrm{fd}~+d_\mathrm{0,max}\cdot \frac{r_\mathrm{w}}{v_\mathrm{max}})\cdot \eta_\mathrm{fd}\cdot\eta_\mathrm{gb}),
		\end{aligned}
	\end{equation}
\end{small}%
\endgroup
where $\gamma_{\mathrm{x}} = \gamma_{\mathrm{fgt}}$ for the FGT, $\gamma_{\mathrm{x}} = \gamma_{\mathrm{min}}$ for the CVT, $T_\mathrm{m,speed}$ is the torque required at the vehicle's top speed, $v_\mathrm{max}$, and can be computed as
\par\nobreak\vspace{-5pt}
\begingroup
\allowdisplaybreaks 
\begin{small}
	\begin{equation}
		\label{eq:Power_vmax}
		T_\mathrm{m,speed} = \frac{P_\mathrm{req}(v_\mathrm{max},0,0)}{v_\mathrm{max}} \cdot r_\mathrm{w}.
	\end{equation}
\end{small}%
\endgroup

%% file: chapters/methodology_7_algorithm.tex
\subsection{Optimization Problem}\label{sec:methodology_optimization_problem}
\ifextendedversion
\else
	We present the optimal design and control problem below. The state variables for both the transmission technologies are given by $x = (E_\mathrm{b},\vartheta_{\mathrm{sft}},\vartheta_{\mathrm{rtr}},\vartheta_{\mathrm{mgt}},\vartheta_{\mathrm{str}},\vartheta_{\mathrm{wdg}},\vartheta_{\mathrm{ewdg}})$. The control and design variables for FGT are $u = P_\mathrm{brk}$ and $p_\mathrm{FGT} = \gamma_\mathrm{fgt}$, respectively. The control variables for CVT are given by $u = (P_\mathrm{brk},\gamma(t))$.
	\begin{prob}[Nonlinear Convex Problem]\label{prob:NLP}
		The minimum-energy design and control strategies are the solution of
		\par\nobreak\vspace{-5pt}
		\begingroup
		\allowdisplaybreaks
		\begin{small}
			\begin{equation*}
				\begin{aligned}
					\min~~&\Delta E_\mathrm{b} = E_{\mathrm{b}}(0) - E_{\mathrm{b}}(\mathrm{T})\\
					\text{s.t. }&\eqref{eq:Power_Balance_FD}-\eqref{eq:Shaft_Losses}, \eqref{eq:Rtr_Loss_Complex}-\eqref{eq:Power_vmax}.
				\end{aligned}
			\end{equation*}
		\end{small}%
		\endgroup
	\end{prob}
 Problem~\ref{prob:NLP} is convex. Yet it cannot be solved by standard convex programming algorithms~\cite{BoydVandenberghe2004}. Nevertheless, we can compute the global optimum by nonlinear programming.
 In order to accelerate convergence, we warm-start it with the solution of a simplified temperature-independent convex quadratically constrained quadratic program (QCQP).
\fi

\subsection{Discussion}
A few comments are in order. 
First, in line with the current practices in high-level design and optimization of automotive powertrains~\cite{VerbruggenSalazarEtAl2019}, we assume constant efficiencies for the FGT, the CVT and neglect the dynamics of the CVT since the transmission modeling is not the aim of this research. We refer readers to~\cite{FahdzyanaSalazarEtAl2020,FahdzyanaSalazarEtAl2021,BorsboomFahdzyanaEtAl2021}, where a more careful analysis of the CVT dynamics is presented.
%
Second, we exclude the gearbox, inverter, and battery temperatures from our thermal models and assume them not to be the limiting factors. However, we can easily extend our framework to account for the temperatures, capturing the full thermal behavior of an EV. Interested readers are directed to~\cite{WeiHofmanEtAl2021} for more information. 
%
Third, convex approximations to the nonlinear EM power losses may result in frequent under- or overestimation of the power losses. This error spreads to the LPTN's neighboring nodes, potentially resulting in diverging temperatures. To mitigate such effects, we identify the LPTN for each driving cycle.
%
Fourth, we consider the average temperatures of each node during fitting and validation of the LPTN to preserve the physical meaning of the LPTN. However, we can emulate hot-spot temperatures by lowering the temperature limits of each component, because replacing average temperatures with hot-spot temperatures may reduce LPTN's accuracy.
%
Finally, we neglect the thermal dynamics of the coolant and assume it to be kept at a constant temperature, $\vartheta_{\mathrm{col}}$, by the radiator. Nevertheless, our results in Section~\ref{sec:results} below show that our models can accurately estimate the temperatures of each component of the EM.

%% file: chapters/results1.tex
\section{Results}\label{sec:results}
\begin{figure}[bt!]	
	\centering
	\includegraphics[trim=0.85cm 0cm 1.4cm 0cm,clip=true, width=\columnwidth]{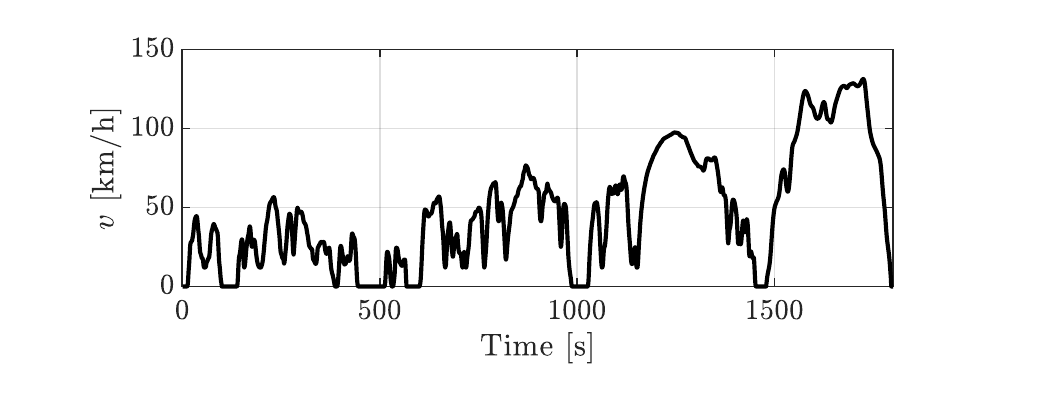}
	\caption{ Worldwide harmonized Light-duty vehicles Test Cycle (WLTC), Class 3. }
	\label{fig:Driving_Cycle}
\end{figure}
\begin{small}
	\begin{table}[t!]
		\centering
		\caption{Parameters.}\scriptsize
		\label{Tab:Parameters}
		\begin{tabular}{l l l l}\toprule
			\textbf{Parameter}   &   \textbf{Symbol}   &   \textbf{Value}   &   \textbf{Units}    \\ \midrule
			\multicolumn{4}{c}{\textit{Vehicle Dynamics \& Transmission}}\\
			Wheel Radius         & $r_{\mathrm{w}}$    & 0.3                & [m]                 \\
			Air drag coefficient & $c_{\mathrm{d}}$    & 0.28               & [-]                 \\
			Frontal Area         & $A_{\mathrm{f}}$    & 2.29               & [m$^\mathrm{2}$]    \\
			Air density          & $\rho_{\mathrm{a}}$ & 1.2041             & [kg/m$^\mathrm{3}$] \\
			Rolling resistance coefficient & $c_{\mathrm{rr}}$ & 0.007      & [-]                 \\
			Gravitational constant & $g$               & 9.81               & [m/s$^\mathrm{2}$]  \\
			Brake fraction       & $r_{\mathrm{b}}$    & 0.65               & [-]                 \\
			Final drive ratio    & $\gamma_{\mathrm{fd,fgt}}$ & 1           & [-]                 \\
								 & $\gamma_{\mathrm{fd,cvt}}$ & 7           & [-]                 \\
			CVT gear ratio limits& $\gamma_{\mathrm{min}}$ & 0.75           & [-]\\ 
								 & $\gamma_{\mathrm{max}}$ & 2.1           & [-]\\ 
			Vehicle base mass    & $m_{\mathrm{0}}$   & 2000                & [kg]                \\
			Gear box mass        & $m_{\mathrm{fgt}}$ & 50                  & [kg]                \\
			& $m_{\mathrm{cvt}}$ & 80                  & [kg]                \\
			Motor to Wheel Efficiency & $\eta_\mathrm{fgt} \cdot \eta_\mathrm{fd}$ & 0.98 & [-]     \\
			& $\eta_\mathrm{cvt} \cdot \eta_\mathrm{fd}$ & 0.96 & [-]     \\
			
			
			\midrule
			\multicolumn{4}{c}{\textit{Thermal Network \& Fan}}\\
			Coolant temperature  & $ \vartheta_{\mathrm{col}}$  & 65                 & $^\circ \mathrm{C}$ \\
			Air temperature gain & $\Delta \vartheta_{\mathrm{a}}$          & 18~\cite{MK_WangJagarwal2014Optimization} & $^\circ \mathrm{C}$ \\
			Specific heat capacity,air & $C_\mathrm{p,air}$ & 1             & kJ/kgK              \\
			Heat exchanger efficiency & $\eta_\mathrm{he}$     & 0.6~\cite{MK_WangJagarwal2014Optimization} & [-]                 \\
			
			\midrule
			\multicolumn{4}{c}{\textit{Battery}}\\
			Battery Capacity     & $E_{\mathrm{b,max}}$ & 37          & [kWh]                \\
			Maximum SoC          & $\zeta_{\mathrm{b,max}}$ & 0.85          & [-]                \\
			Minimum SoC          & $\zeta_{\mathrm{b,min}}$ & 0.15          & [-]                \\
			
			\midrule
			\multicolumn{4}{c}{\textit{Performance Requirements}}\\
			Starting Gradient    & $\alpha_{\mathrm{start}}$ & 0.2          & [-]                \\
			Top Speed            & $v_{\mathrm{top}}$        & 135          & [kmph]             \\
			Acceleration Time    & $t_{\mathrm{acc}}$        & 15           & [s]             \\
			Acceleration Speed   & $v_{\mathrm{acc}}$        & 100          & [kmph]             \\
			
			\bottomrule
		\end{tabular}
	\end{table}
\end{small}%
\begin{small}
	\begin{table}[t!]
		\centering
		\caption{Motor Specifications.}\scriptsize
		\label{Tab:Motor_Specs}
		\begin{tabular}{l | l r | l r | l r | l r}\toprule
			& \multicolumn{2}{c|}{Motor 1}& \multicolumn{2}{c|}{Motor 2}& \multicolumn{2}{c|}{Motor 3}& \multicolumn{2}{c}{Motor 4} \\
			
			\midrule
			$m_\mathrm{m}$ [kg]				& \multicolumn{2}{c|}{50.66}   & \multicolumn{2}{c|}{42.04}   & \multicolumn{2}{c|}{24.58}   & \multicolumn{2}{c}{24.58}   \\
			$T_\mathrm{m,max}$	[Nm]		& \multicolumn{2}{c|}{287}     & \multicolumn{2}{c|}{228}     & \multicolumn{2}{c|}{145}     & \multicolumn{2}{c}{145}     \\
			$P_\mathrm{m,max}$ 	[kW]		& \multicolumn{2}{c|}{134}     & \multicolumn{2}{c|}{132}     & \multicolumn{2}{c|}{112}     & \multicolumn{2}{c}{112}     \\
			$\omega_\mathrm{m,max}$ [rad/s]	& \multicolumn{8}{c}{1047} \\
			$\omega_\mathrm{m,b}$ [rad/s]	& \multicolumn{2}{c|}{419}     & \multicolumn{2}{c|}{550}     & \multicolumn{2}{c|}{733}     & \multicolumn{2}{c}{733}     \\
			$\dot{m}_\mathrm{col}$ [l/min]	& \multicolumn{2}{c|}{6.5}     & \multicolumn{2}{c|}{5.2}     & \multicolumn{2}{c|}{0.2}     & \multicolumn{2}{c}{0.025}     \\
			\bottomrule
		\end{tabular}
	\end{table}
\end{small}%
This section presents the numerical results obtained when we apply the framework presented in Section~\ref{sec:methodology} above to optimize the powertrain design and control strategies of a compact family car. In line with current practices for optimizing the design and control of hybrid electric vehicles~\cite{GuzzellaSciarretta2007}, we consider two different driving cycles: World harmonized Light-vehicles Test Cycle (WLTC) Class 3 and a custom cycle obtained by repeating the WLTC Class 3 twice, further referred to as WLTCx2. Fig.~\ref{fig:Driving_Cycle} shows the velocity profile of the WLTC. In addition, we use the WLTCx2 cycle to simulate extended driving scenarios, thereby thermally stress testing the EM. We optimize the control strategies for an electric powertrain equipped with four motors as shown in Fig.~\ref{fig:Motor_Cross_Section}, two transmissions (FGT and CVT) simulated on two drive cycles (WLTC and WLTCx2), resulting in 16 unique combinations.

Table~\ref{Tab:Parameters} shows the vehicle parameters required to obtain the numerical results presented in this section and Table~\ref{Tab:Motor_Specs} shows the specifications of the four EMs. In line with~\cite{VerbruggenSalazarEtAl2019}, we discretize the optimization problem with a sampling time of \unit[1]{s} using trapezoidal integration in order to avoid numerical instabilities stemming from the LPTN~\cite{LocatelloKondaEtAl2020}. We parse the problem in CasADi~\cite{AnderssonGillisEtAl2019} and solve it with the nonlinear solver IPOPT~\cite{WachterBiegler2006}. Overall, It takes about \unit[50]{s} to parse and \unit[100]{s} to converge for one motor-transmission-drive cycle combination when using a machine with Intel\textregistered~Core\texttrademark~i7-9750H CPU and \unit[16]{GB} of RAM.

The remainder of the results are presented as follows: First, we validate the accuracy of our results by comparing it with the data obtained from the high-fidelity simulation software Motor-CAD. Second, we present a case study comparing the difference between an electric powertrain equipped with an FGT and a CVT. Third, we visualize a case in which the motor fails thermally. Fourth, we compare the power distribution between fan and friction brakes to show that it is more efficient to control the EM temperatures with a fan. Finally, we compare the optimization results for all the motor-transmission-drive cycle combinations.

\subsection{Validation}\label{res:validation}
\begin{figure}[bt!]	
	\centering
	\includegraphics[trim=1.85cm 2.25cm 2cm 11cm,clip=true, width=\columnwidth]{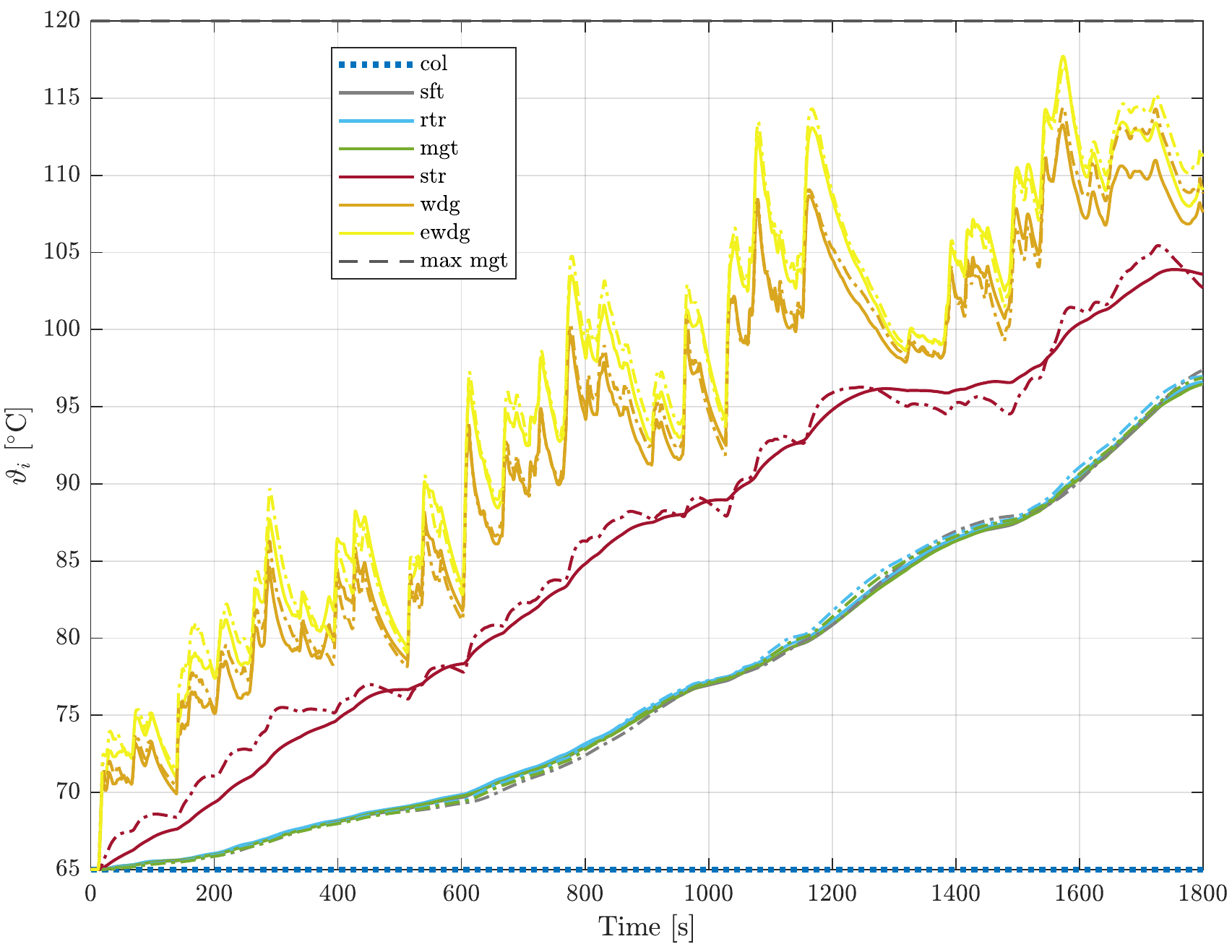}
	\caption{ The optimization results for a vehicle equipped with Motor~3, an FGT, and simulated over the WLTC cycle. The graph shows the EM components' temperature, their limits and a validation with high-fidelity simulation software Motor-CAD. The results are presented with solid lines and simulation data from Motor-CAD is presented with dash-dotted lines.}
	\label{fig:AR751_SRT_WLTP}
\end{figure}%
Using Algorithm~\ref{alg:algorithm}, we optimize the control strategies for a powertrain equipped with Motor 3, an FGT and simulated over the WLTC. Fig.~\ref{fig:AR751_SRT_WLTP} shows the optimization results and validation with the high-fidelity simulation software Motor-CAD. We can observe that our models closely reproduce the thermal behavior from Motor-CAD, resulting in a cumulative drift below 1$^\circ$C for all the components except the end-windings, whose temperature drifts by 2$^\circ$C.

\subsection{Case Study: Comparison between an FGT and a CVT}\label{res:FGTvsCVT}
\begin{figure}[t!]	
	\centering
	\includegraphics[trim=1.85cm 2.25cm 2cm 2.25cm,clip=true, width=\columnwidth]{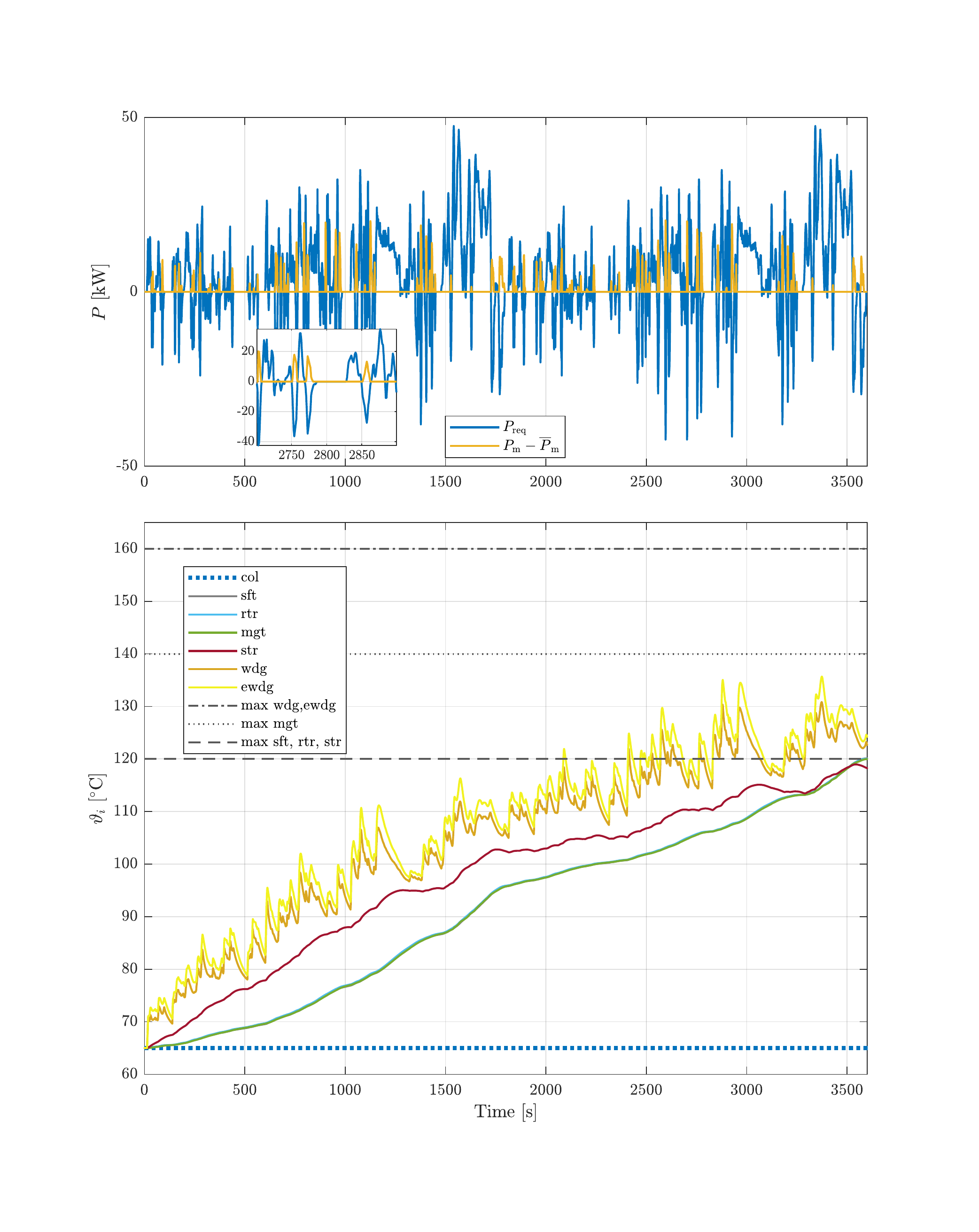}
	\caption{ The resulting powers, temperatures for an electric powertrain equipped with Motor 3, FGT and simulated over the WLTCx2 cycle. }
	\label{fig:AR751_SRT_WLTP_Rep2}
\end{figure}%
\begin{figure}[t!]	
	\centering
	\includegraphics[trim=1.85cm 2.25cm 2cm 2.25cm,clip=true, width=\columnwidth]{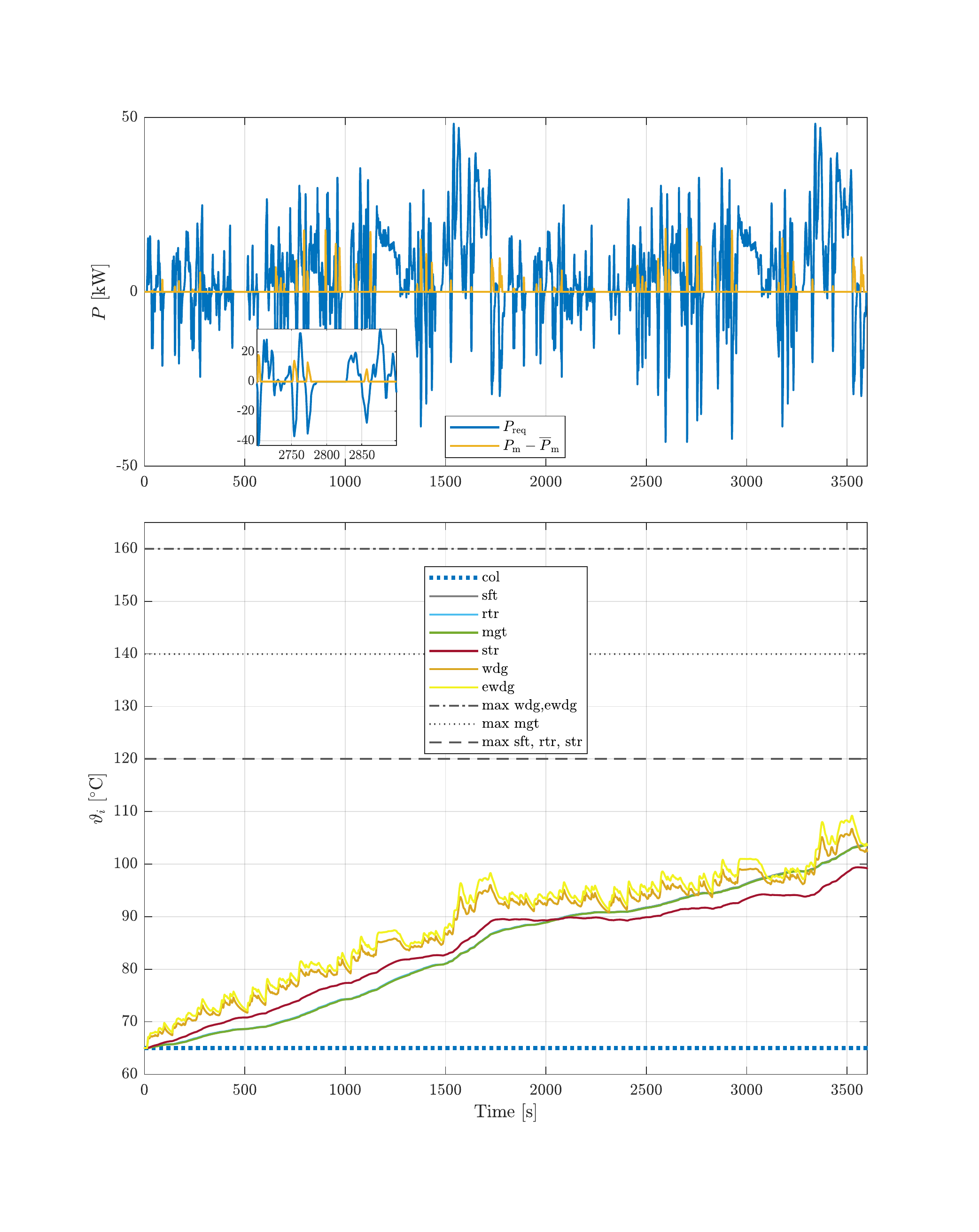}
	\caption{ The resulting powers, temperatures for an electric powertrain equipped with Motor 3, CVT and simulated over the WLTCx2 cycle. }
	\label{fig:AR751_CVT_WLTP_Rep2}
\end{figure}%
Fig.~\ref{fig:AR751_SRT_WLTP_Rep2} showcases the optimization results for a powertrain with Motor 3, an FGT and simulated over the WLTCx2 cycle. Whilst Fig.~\ref{fig:AR751_CVT_WLTP_Rep2} presents the optimization results for a powertrain equipped with Motor 3, a CVT and simulated over the WLTCx2 cycle. The required power, $P_\mathrm{req}$, and the regenerative power lost to friction brakes, $P_{\mathrm{m}} - \overline{P}_{\mathrm{m}}$, are plotted in the top subplot, and the evolution of the components' temperatures is shown in the lower subplot. We can observe that the EM's components reach higher temperatures when compared to Section~\ref{res:validation} because of extended driving. In addition, the powertrain equipped with an FGT reaches the thermal boundaries, whilst the powertrain equipped with a CVT can operate on the optimal operating line whenever possible, thereby improving the motor efficiency and consequently having overall lower temperatures. We can also observe from the power plots that a higher amount of regenerative power is diverted to mechanical brakes in the case of an FGT than a CVT to keep the EM temperatures within their limits. This reduces the range achievable by the vehicle as shown in Fig.~\ref{fig:Bar_Graph}. Also, the permanent magnets are the limiters during extended driving scenarios. Since a lot of potential regenerative braking energy is lost to friction brakes in the case of an FGT, its range is reduced from \unit[165.4]{km} for the thermally unconstrained case to \unit[156.4]{km} for the thermally constrained case which is a substantial decrease of 5.4\%. We can conclude that the CVT is preferred for the thermally constrained case and has the potential to further downsize the EM.

\subsection{Case Study: Thermal Failure}
\begin{figure}[t!]	
	\centering
	\includegraphics[trim=1.85cm 2.25cm 2cm 2.25cm,clip=true, width=\columnwidth]{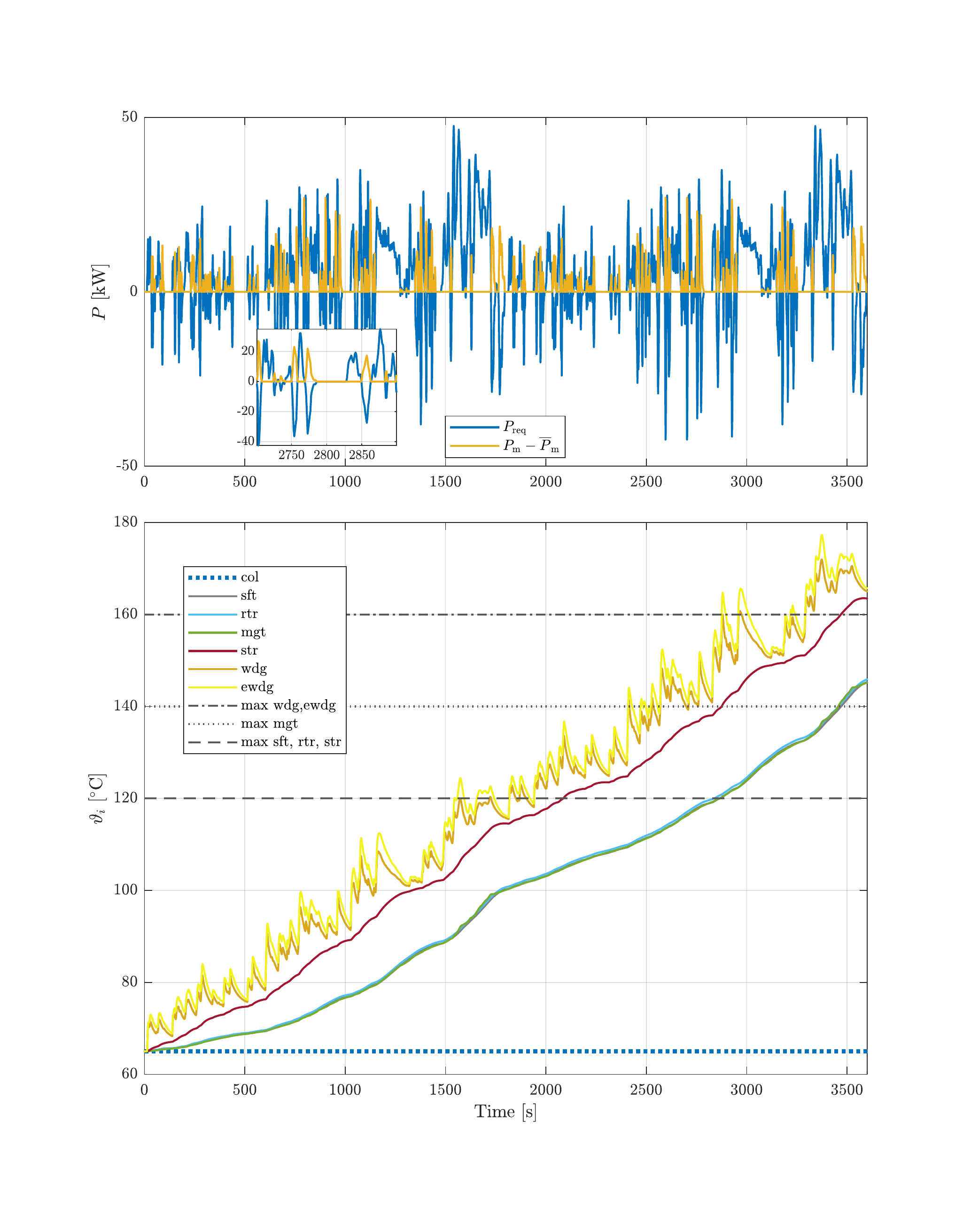}
	\caption{ The resulting powers, temperatures for an electric powertrain equipped with Motor 4, FGT and simulated over the WLTCx2 cycle. }
	\label{fig:AR752_SRT_WLTP_Rep2}
\end{figure}
In this section, we showcase a scenario in which the motor fails thermally. In order to do so, we consider a powertrain equipped with Motor 4 and an FGT simulated over the WLTCx2 cycle. In addition, we deactivate the regeneration potential of the EM.
We can observe from the optimization results in Fig.~\ref{fig:AR752_SRT_WLTP_Rep2} that the thermal limits are breached even though regenerative braking is not allowed, which suggests that the motor is not suitable for operation in this scenario. Furthermore, we can conclude that the regenerative braking control has its limitations and can only reduce the EM's temperatures to a certain extent. Moreover, we could increase the coolant flow rate to reject more heat and cool down the EM allowing it to operate within its thermal limits. For example, Motor 4 with a lower coolant flow rate fails thermally, while Motor 3 with a higher coolant flow rate can operate over the WLTCx2 without breaching any limits.

\subsection{Case Study: Fan Power vs Recuperation Power} \label{res:FanvsBrake}
\begin{figure}[t!]	
	\centering
	\includegraphics[trim=0.75cm 0.5cm 1.5cm 0.95cm,clip=true, width=\columnwidth]{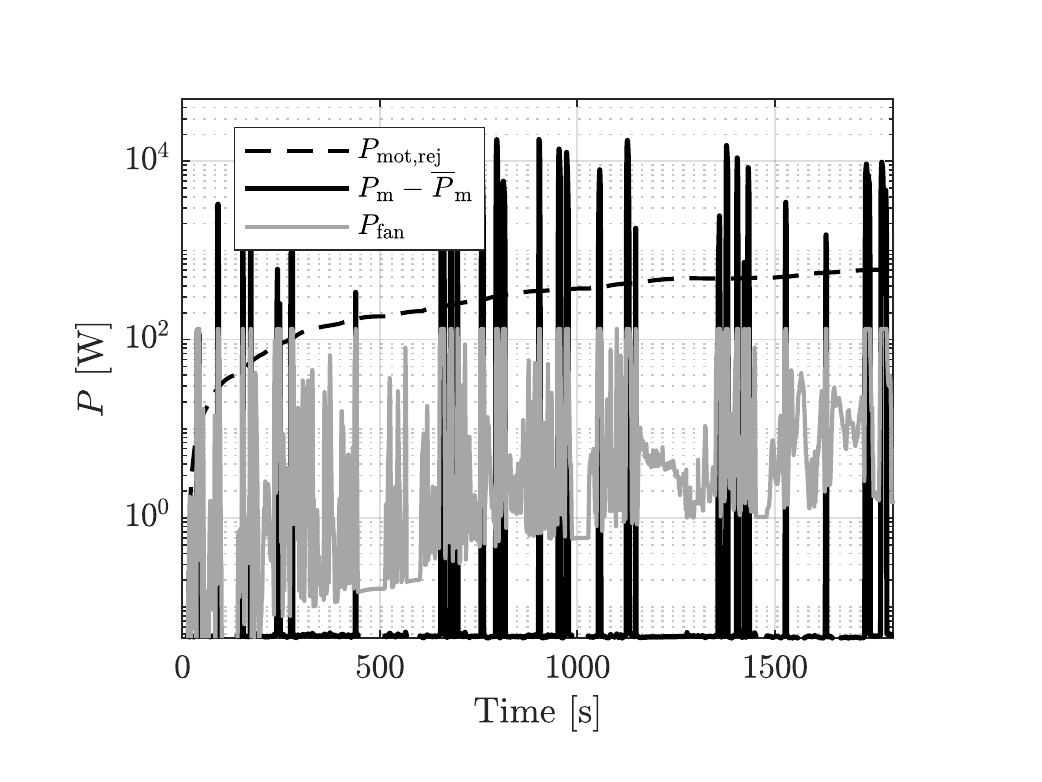}
	\caption{ Comparison of the heat rejected by the motor, the fan power, and the recuperative power lost to friction brakes. }
	\label{fig:FanPowervsBrakePower}
\end{figure}%
We consider the scenario from Section~\ref{res:validation} with Motor 3, an FGT and simulated over WLTC.  Fig.~\ref{fig:FanPowervsBrakePower} shows the heat rejected by the motor, $P_\mathrm{mot,rej}$, potential regenerative motor power lost to mechanical brakes, $P_{\mathrm{m}} - \overline{P}_{\mathrm{m}}$, and the fan power ($P_\mathrm{fan}$). 
We can observe that operating a fan is preferred over losing energy to mechanical brakes because the fan power is less expensive, as the fan consumes less energy to remove heat from the EM than the friction brakes.

\subsection{Comparison of Results}\label{res:ComparisionResults}
\begin{tiny}
	\begin{table}[t!]
		\centering
		\caption{Optimization Results.}\scriptsize
		\label{Tab:Simulation results}
		\begin{tabular}{c | l r | l r | l r | l r}\toprule
			& \multicolumn{2}{c|}{Motor 1}& \multicolumn{2}{c|}{Motor 2}& \multicolumn{2}{c|}{Motor 3}& \multicolumn{2}{c}{Motor 4} \\
			& FGT          & CVT          & FGT          & CVT          & FGT          & CVT          & FGT          & CVT          \\
			
			\midrule
			\multicolumn{9}{l}{\textbf{WLTC}}\\
			$\gamma$						& 5.34         & -             & 6.39         & -            & 8.38         & -            & 8.38        & - \\
			$\overline{E}_\mathrm{b}$			& 154.8        & 154.4         & 155.7        & 154.9        & 156.8        & 153.6        & 156.1       & 153.2\\
			$D$								& 167.4        & 167.9         & 166.5        & 167.3        & 165.4        & 168.7        & 166.0       & 169.2\\
			
			\midrule
			\multicolumn{9}{l}{\textbf{WLTCx2}}\\
			$\gamma$						& 5.42         &  -            & 6.48         & -            & 8.38         & -              &             &\\
			$\overline{E}_\mathrm{b}$			& 155.0        &  154.5        & 156.0        & 155.1        & 165.7        & 154.0          & N/A         & N/A \\
			$D$								& 167.2        &  167.8        & 166.1        & 167.1        & 156.4        & 168.2          &             & \\
			
			\bottomrule
		\end{tabular}%
	\end{table}%
\end{tiny}%
\begin{figure}[t!]	
	\centering
	\includegraphics[trim=0.75cm 0.5cm 0.75cm 0.5cm,clip=true, width=0.9\columnwidth]{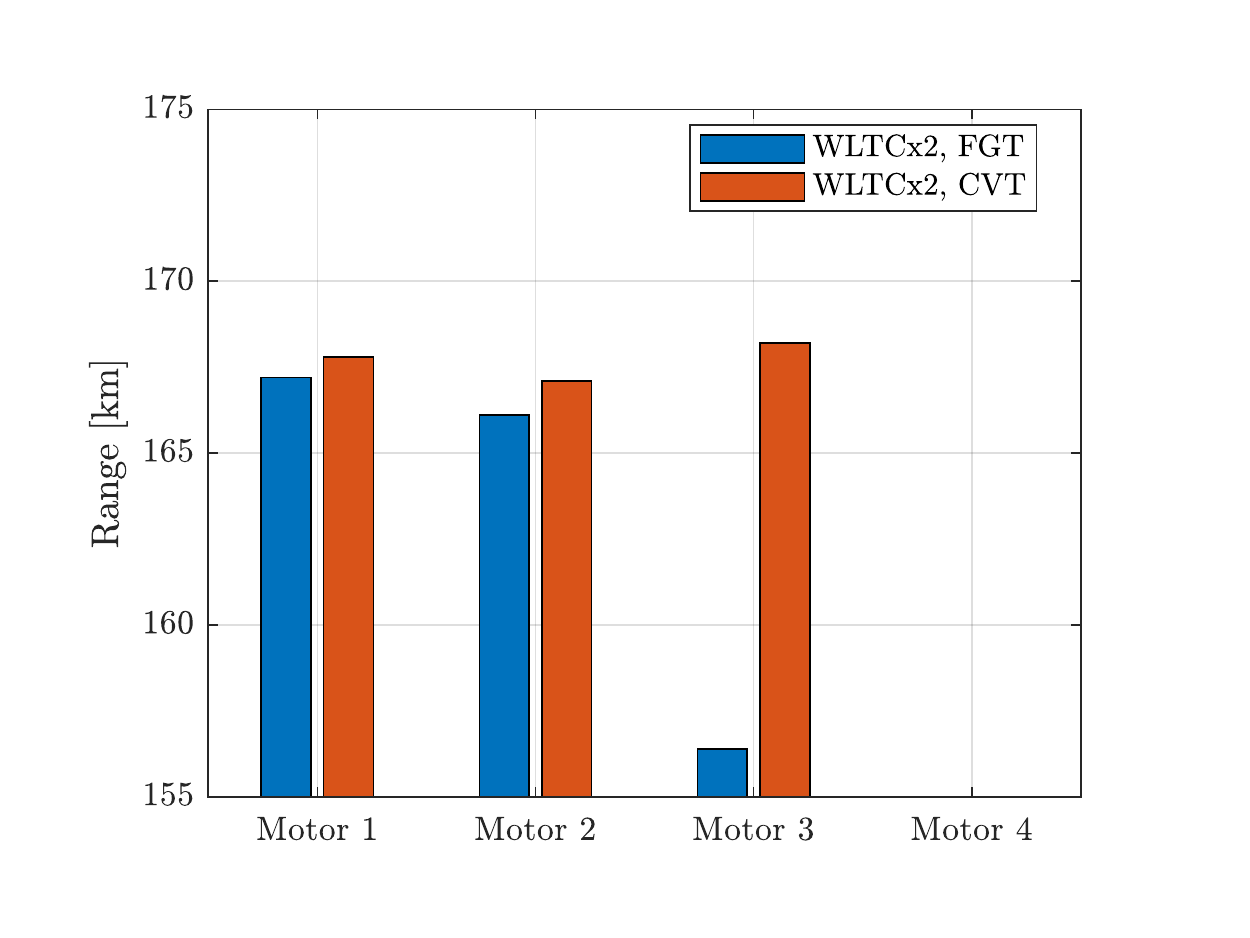}
	\caption{ Comparison of Range for different motor-transmission combinations simulated over WLTCx2. }
	\label{fig:Bar_Graph}
\end{figure}%
This section presents the optimization results for all the motor, transmission, and drive cycle combinations. Table~\ref{Tab:Simulation results} summarizes the results where $D$ is the achievable range in \unit{km}, $\overline{E}_\mathrm{b}$ is the rate of energy consumption in \unit{Wh/km}. Since Motor 4 fails thermally when simulated over the WLTCx2, we do not present the optimization results for these scenarios. Fig.~\ref{fig:Operating_Points} shows the operating points of the WLTCx2 for the four EMs. We can observe that the CVT operates at lower speeds than the FGT because the speed contributes to higher losses than torques, as shown in Fig.~\ref{fig:Ploss_Data_Model}, making the CVT operate at lower speeds. Fig.~\ref{fig:Bar_Graph} compares the achievable range, $D$, for all the motors simulated over WLTCx2. Despite the CVT being heavy and less efficient, the CVT equipped powertrains have longer ranges because of their efficient operation. This trend is prominent for Motor 3, where the FGT has a shorter range because of its thermally constrained operation. As shown in Fig.~\ref{fig:AR751_CVT_WLTP_Rep2}, the CVT equipped powertrains operate at lower temperatures than the FGT equipped powertrains. Finally, we must note that the maximum motor power of Motor 3 is 16.4\% lower than Motor 1 and the coolant flow rate is 97\% lower. This result shows the downsizing potential associated with the effective use of the peak performance envelope of the EM.

%% file: chapters/results.tex
\section{Results}\label{sec:results}
	\begin{small}
		\begin{table}[t!]
			\centering
			\caption{Simulation Parameters.}\scriptsize
			\label{Tab:Parameters}
			\begin{tabular}{l l l l}\toprule
				\textbf{Parameter}   &   \textbf{Symbol}   &   \textbf{Value}   &   \textbf{Units}    \\ \midrule
				\multicolumn{4}{c}{\textit{Vehicle Dynamics \& Transmission}}\\
				Wheel Radius         & $r_{\mathrm{w}}$    & 0.3                & [m]                 \\
				Air drag coefficient & $c_{\mathrm{d}}$    & 0.28               & [-]                 \\
				Frontal Area         & $A_{\mathrm{f}}$    & 2.29               & [m$^\mathrm{2}$]    \\
				Air density          & $\rho_{\mathrm{a}}$ & 1.2041             & [kg/m$^\mathrm{3}$] \\
				Rolling resistance coefficient & $c_{\mathrm{rr}}$ & 0.007      & [-]                 \\
				Gravitational constant & $g$               & 9.81               & [m/s$^\mathrm{2}$]  \\
				Brake fraction       & $r_{\mathrm{b}}$    & 0.65               & [-]                 \\
				Final drive ratio    & $\gamma_{\mathrm{fd,fgt}}$ & 1           & [-]                 \\
									 & $\gamma_{\mathrm{fd,cvt}}$ & 7           & [-]                 \\
				CVT gear ratio limits& $\gamma_{\mathrm{min}}$ & 0.75           & [-]\\ 
									 & $\gamma_{\mathrm{max}}$ & 2.1           & [-]\\ 
				Vehicle base mass    & $m_{\mathrm{0}}$   & 2000                & [kg]                \\
				Gear box mass        & $m_{\mathrm{fgt}}$ & 50                  & [kg]                \\
				& $m_{\mathrm{cvt}}$ & 80                  & [kg]                \\
				Motor to Wheel Efficiency & $\eta_\mathrm{fgt} \cdot \eta_\mathrm{fd}$ & 0.98 & [-]     \\
				& $\eta_\mathrm{cvt} \cdot \eta_\mathrm{fd}$ & 0.96 & [-]     \\
				
				
				\midrule
				\multicolumn{4}{c}{\textit{Thermal Network \& Fan}}\\
				Coolant temperature  & $ \vartheta_{\mathrm{col}}$  & 65                 & $^\circ \mathrm{C}$ \\
				Air temperature gain & $\Delta \vartheta_{\mathrm{a}}$          & 18~\cite{WangJagarwalEtAl2015} & $^\circ \mathrm{C}$ \\
				Specific heat capacity,air & $C_\mathrm{p,air}$ & 1             & kJ/kgK              \\
				Heat exchanger efficiency & $\eta_\mathrm{he}$     & 0.6~\cite{WangJagarwalEtAl2015} & [-]                 \\
				
				\midrule
				\multicolumn{4}{c}{\textit{Battery}}\\
				Battery Capacity     & $E_{\mathrm{b,max}}$ & 37          & [kWh]                \\
				Maximum SoC          & $\zeta_{\mathrm{b,max}}$ & 0.85          & [-]                \\
				Minimum SoC          & $\zeta_{\mathrm{b,min}}$ & 0.15          & [-]                \\
				
				\midrule
				\multicolumn{4}{c}{\textit{Performance Requirements}}\\
				Starting Gradient    & $\alpha_{\mathrm{start}}$ & 0.2          & [-]                \\
				Top Speed            & $v_{\mathrm{top}}$        & 135          & [kmph]             \\
				Acceleration Time    & $t_{\mathrm{acc}}$        & 15           & [s]             \\
				Acceleration Speed   & $v_{\mathrm{acc}}$        & 100          & [kmph]             \\
				
				\bottomrule
			\end{tabular}
		\end{table}
	\end{small}%
%
	\begin{small}
	\begin{table}[t!]
		\centering
		\caption{Electric Motor Specifications.}\scriptsize
		\label{Tab:Motor_Specs}
		\begin{tabular}{l | l r | l r | l r }\toprule
			& \multicolumn{2}{c|}{Motor 1}& \multicolumn{2}{c|}{Motor 2}& \multicolumn{2}{c}{Motor 3}\\
			
			\midrule
			$m_\mathrm{m}$ [kg]				& \multicolumn{2}{c|}{50.66}   & 42.04 & (-17.0\%)         & 24.58 & (-51.5\%)   \\
			$T_\mathrm{m,max}$	[Nm]		& \multicolumn{2}{c|}{287}     & 228   & (-20.6\%)        & 145   & (-49.5\%)     \\
			$P_\mathrm{m,max}$ 	[kW]		& \multicolumn{2}{c|}{134}     & 132   & (-1.5\%)         & 112   & (-16.4\%)     \\
			$\omega_\mathrm{m,max}$ [rad/s]	& \multicolumn{6}{c}{1047}                                                                   \\
			$\omega_\mathrm{m,b}$ [rad/s]	& \multicolumn{2}{c|}{419}     & \multicolumn{2}{c|}{550}  & \multicolumn{2}{c}{733}     \\
			$\dot{m}_\mathrm{col}$ [l/min]	& \multicolumn{2}{c|}{6.5}     & 5.2   & (-20.0\%)        & 0.2   & (-96.9\%)     \\
			\bottomrule
		\end{tabular}
	\end{table}
\end{small}%
\begin{table}[t!]
\caption{Thermal limits of the nodes of the thermal network.}
\centering
\begin{tabular}{l c l c}
	\toprule
	Component  & $\vartheta_{i,\mathrm{max}}$ [\unit{$^ \circ$C}] & Component  & $\vartheta_{i,\mathrm{max}}$ [\unit{$^ \circ$C}]\\
	\midrule
	Shaft          & 140 & Permanent Magnets        & 120 \\
	Rotor          & 140 & Stator                   & 140 \\
	Windings       & 160 & End-windings             & 160 \\
	\bottomrule
\end{tabular}
\label{tab:ThLim}
\end{table}%
This section presents the numerical results obtained when we apply the framework presented in Section~\ref{sec:methodology} to optimize the powertrain design and control strategies of a compact family car. In line with current practices for hybrid electric vehicles~\cite{GuzzellaSciarretta2007}, we optimize the powertrain design and control for given driving cycles: the World harmonized Light-vehicles Test Cycle (WLTC) Class 3 and a custom cycle obtained by repeating the WLTC Class 3 twice, further referred to as WLTCx2.
In addition, we use the WLTCx2 to simulate extended driving scenarios, thereby thermally stress testing the EM. 
\ifextendedversion
\else
	We optimize the control strategies for an electric powertrain equipped on of the three motors detailed in Table~\ref{Tab:Motor_Specs}, an FGT or a CVT, and on two drive cycles (WLTC and WLTCx2), resulting in 12 unique combinations.
\fi

\ifextendedversion
\else
	Table~\ref{Tab:Parameters} shows the vehicle parameters used to obtain the numerical results presented in this section, and the motor specifications are summarized in Table~\ref{Tab:Motor_Specs} and fitted from Motor-CAD data~\cite{MotorCAD}. 
	Motor 1, shown in the top subplot of Fig.~\ref{fig:ThermalNetwork}, is based on the 2011 Nissan Leaf's EM, whilst Motors 2 and 3 are created by reducing its coolant flow rates, and scaling its dimensions, resulting in lower peak power and torque levels.
\fi
Finally, Table~\ref{tab:ThLim} summarizes the maximum temperatures of all the nodes.
In line with~\cite{VerbruggenSalazarEtAl2019}, we discretize the optimization problem with a sampling time of \unit[1]{s} using trapezoidal integration in order to avoid numerical instabilities stemming from the LPTN~\cite{LocatelloKondaEtAl2020}. We parse the problem in CasADi~\cite{AnderssonGillisEtAl2019} and solve it with the nonlinear solver IPOPT~\cite{WachterBiegler2006}. Overall, It takes about \unit[50]{s} to parse and \unit[100]{s} to converge for one motor-transmission-drive cycle combination when using a computer with Intel\textregistered~Core\texttrademark~i7-9750H CPU and \unit[16]{GB} of RAM.

The remainder of the results are presented as follows: First, we validate the accuracy of our models by comparing them with the high-fidelity simulation software Motor-CAD~\cite{MotorCAD}. Second, we present a case study comparing an electric powertrain equipped with an FGT and a CVT. 
Finally, we compare the optimization results for all the motor-transmission-drive cycle combinations.

\subsection{Validation}\label{res:validation}
\begin{figure}[t!]	
	\centering
	\includegraphics[trim=1.45cm 0.36cm 1.76cm 0.9cm,clip=true, width=\columnwidth]{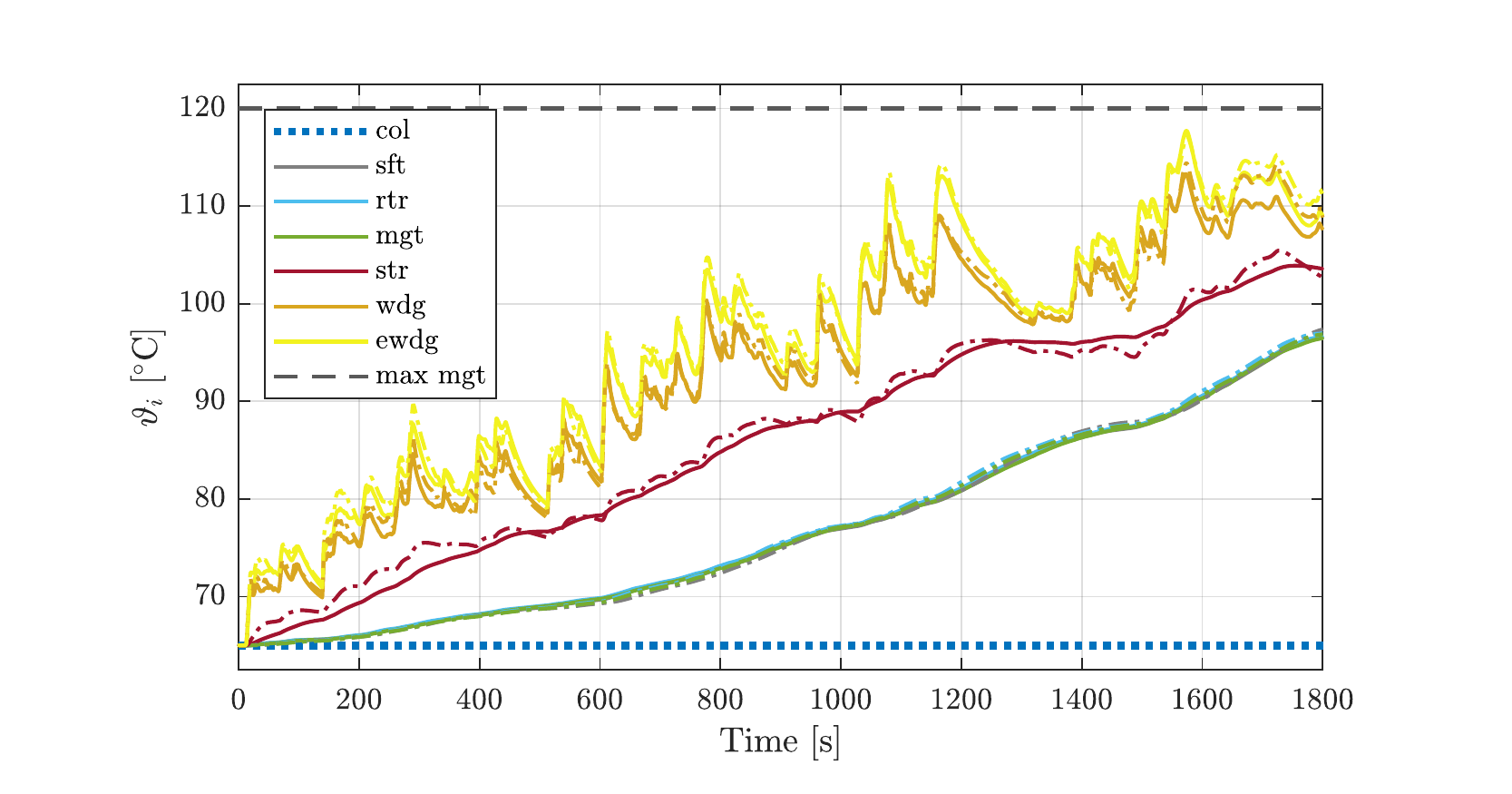}
	\caption{Optimization results for a vehicle equipped with Motor~3, an FGT, and simulated over the WLTC. The plot shows the EM components' temperature (solid), their limits (dashed) and a validation with the high-fidelity simulation software Motor-CAD (dash-dotted).}
	\label{fig:AR751_SRT_WLTP}
\end{figure}%
\ifextendedversion
\else
	In order to validate our models, we solve the optimal control problem of a powertrain equipped with Motor 3 and an FGT on the WLTC. 
\fi
Fig.~\ref{fig:AR751_SRT_WLTP} shows that our models closely reproduce the thermal behavior from Motor-CAD, resulting in a cumulative drift below 1$^\circ$C for all the components except the end-windings, whose temperature drifts by 2$^\circ$C.

\subsection{Case Study: Comparison between an FGT and a CVT}\label{res:FGTvsCVT}
\begin{figure}[t!]	
	\centering
	\includegraphics[trim=1.45cm 0.36cm 1.76cm 0.9cm,clip=true, width=\columnwidth]{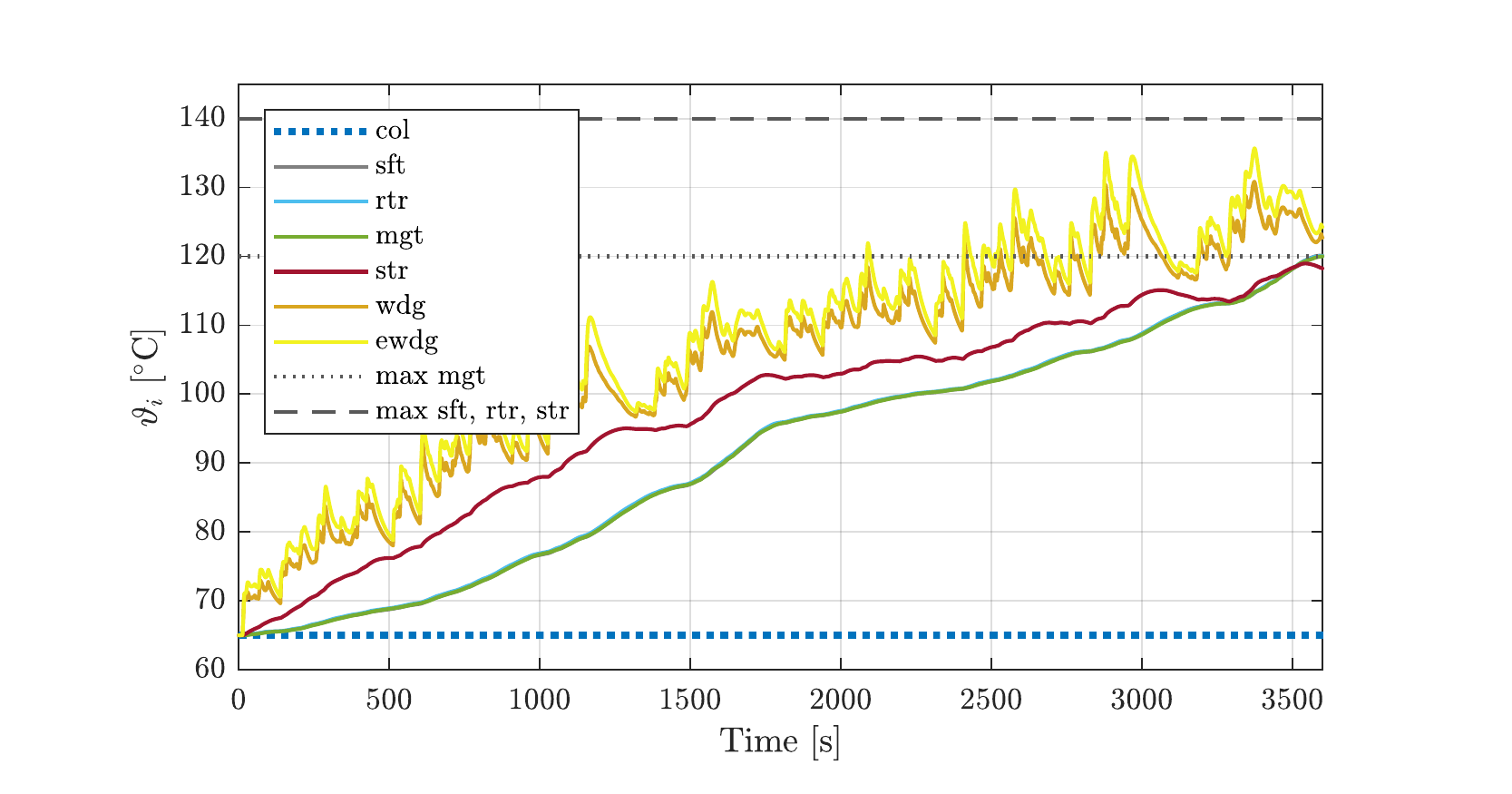}
	\caption{ The resulting temperatures for an electric powertrain equipped with Motor 3, an FGT and simulated over the WLTCx2 cycle. }
	\label{fig:AR751_SRT_WLTP_Rep2}
\end{figure}%
\begin{figure}[t!]	
	\centering
	\includegraphics[trim=1.45cm 0.36cm 1.76cm 0.9cm,clip=true, width=\columnwidth]{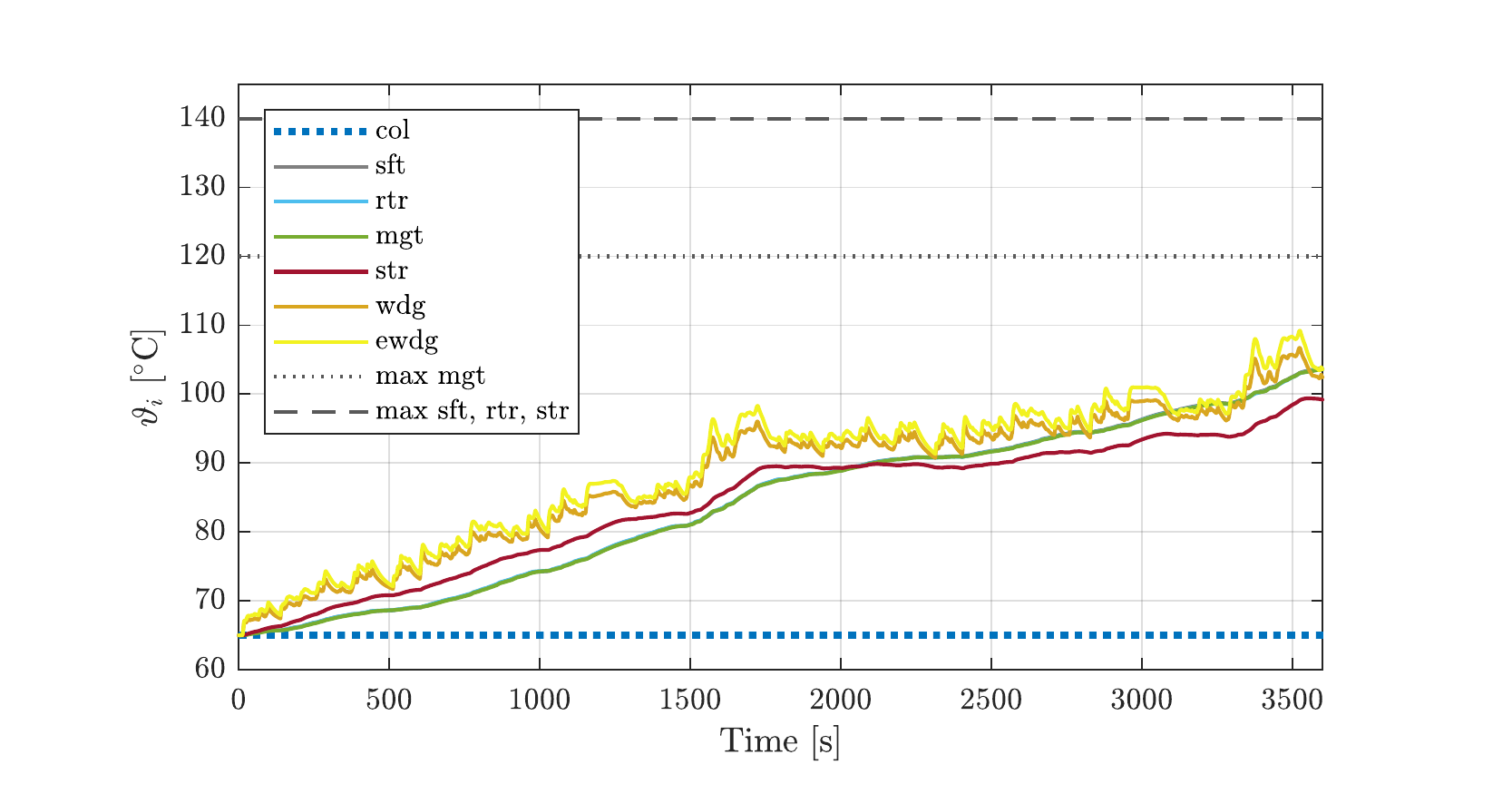}
	\caption{ The resulting temperatures for an electric powertrain equipped with Motor 3, a CVT and simulated over the WLTCx2 cycle. }
	\label{fig:AR751_CVT_WLTP_Rep2}
\end{figure}%
Fig.~\ref{fig:AR751_SRT_WLTP_Rep2} showcases the optimization results for a powertrain with Motor 3, an FGT and simulated over the WLTCx2. Whilst Fig.~\ref{fig:AR751_CVT_WLTP_Rep2} presents the optimization results for a powertrain equipped with Motor 3, a CVT and simulated over the WLTCx2 cycle. 
The evolution of the components' temperatures is shown in the plots. We can observe that the EM's components reach higher temperatures when compared to Section~\ref{res:validation} because of extended driving. In addition, the powertrain equipped with an FGT reaches the thermal boundaries of the magnets, whilst the powertrain equipped with a CVT can operate on the optimal operating line whenever possible, thereby improving the motor efficiency and consequently having overall lower temperatures.
Thereby, regenerative braking was always favored at the cost of a higher fan operation.
Moreover, a larger amount of regenerative power is diverted to the mechanical brakes in the case of an FGT, as it is forced to keep the EM temperatures within their limits, resulting into the CVT achieving a lower energy consumption. 

\subsection{Comparison of Results}\label{res:ComparisionResults}
\begin{figure}[t!]	
	\centering
	\includegraphics[trim=1.2cm 0.38cm 1.7cm 0.3cm,clip=true,  width=\columnwidth]{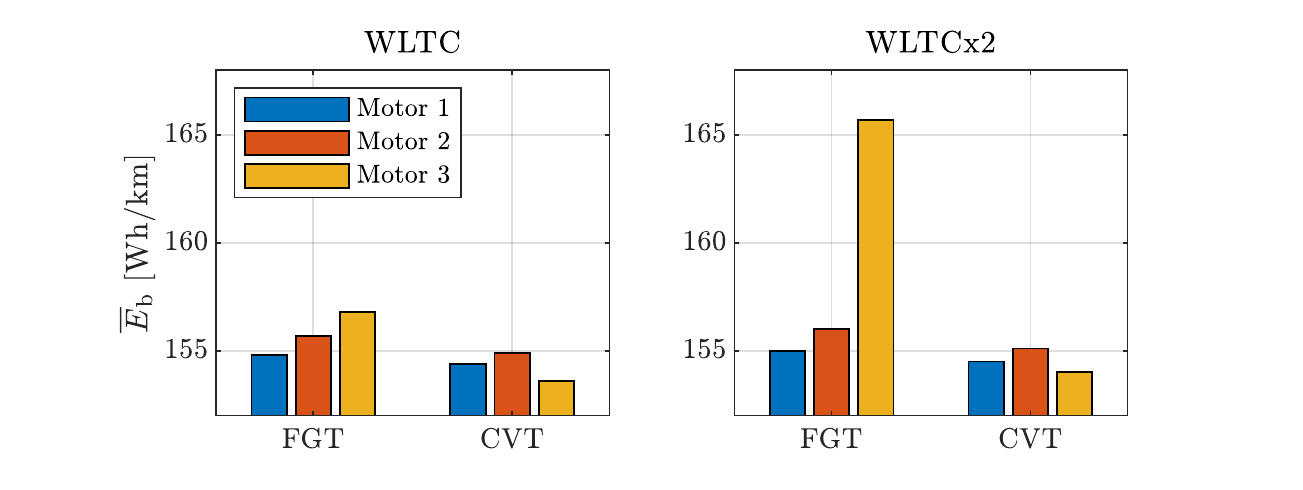}
	\caption{Comparison of the energy consumption rates for different motor-transmission combinations.}
	\label{fig:BarGraph1}
\end{figure}%
Fig.~\ref{fig:BarGraph1} presents the optimization results for all the motor, transmission, and drive cycle combinations, where $\overline{E}_\mathrm{b}$ is the energy consumption per distance traveled $D$, and mathematically defined as
\par\nobreak\vspace{-5pt}
\begingroup
\allowdisplaybreaks 
\begin{small}
	\begin{equation}
		\label{eq:Eb_bar}
		\overline{E}_\mathrm{b} = \frac{\Delta E_\mathrm{b}}{D}.
	\end{equation}
\end{small}%
\endgroup

Despite the CVT being heavier and with a lower mechanical efficiency, the CVT-equipped powertrains have lower energy consumption because their more efficient EM operation results in fewer losses, whilst enabling more regenerative braking.
This result shows the downsizing potential associated with the effective use of the EM's peak performance envelope stemming from CVTs.

%% file: chapters/conclusion.tex
\section{Conclusion}\label{sec:conclusion}
In this paper, we investigated methods for jointly optimizing the design and control strategies of an all-electric powertrain in terms of energy consumption while explicitly accounting for the thermal behavior of the electric motor (EM). Specifically, we explored the trade-off between maximizing regenerative braking and minimizing radiator usage to maintain the temperature of the EM within its limits. To this end, we derived convex models of the vehicle components, determined the performance requirements, formulated a convex optimization problem and applied our methods to design an electric compact family car. 
Our validation with Motor-CAD showed that our models accurately captured the thermal behavior of the EM, and the permanent magnets' temperature is the limiting factor during extended driving.
Furthermore, our results revealed that continuously variable transmission (CVT) equipped powertrains can operate at lower EM temperatures w.r.t.\ fixed gear transmission equipped powertrains, as the CVT keeps the EM on the maximum efficiency line whenever possible, resulting in lower losses and consequently less heat generation, hence enabling more regenerative braking. Finally, we could decrease the maximum power of the EM by 16\%, highlighting the importance of considering the thermal behavior when designing an electric powertrain.


This research opens the field for the following extensions: First, we would like to include the thermal dynamics of the inverter, transmission, and battery to exhaustively capture the thermal behavior of an electric powertrain. 
Second, we would like to control the active aerodynamic elements to regulate the airflow over the radiator for cooling purposes.
Third, we would like to extend the methods for real time control of electric vehicles.


%% file: chapters/appendix.tex
\subsection{EM Identification} \label{app:Motor_Limits}
\begin{figure}[H]	
	\centering
	\includegraphics[trim=1.15cm 0.25cm 1.6cm 0.65cm,clip=true, width=\columnwidth]{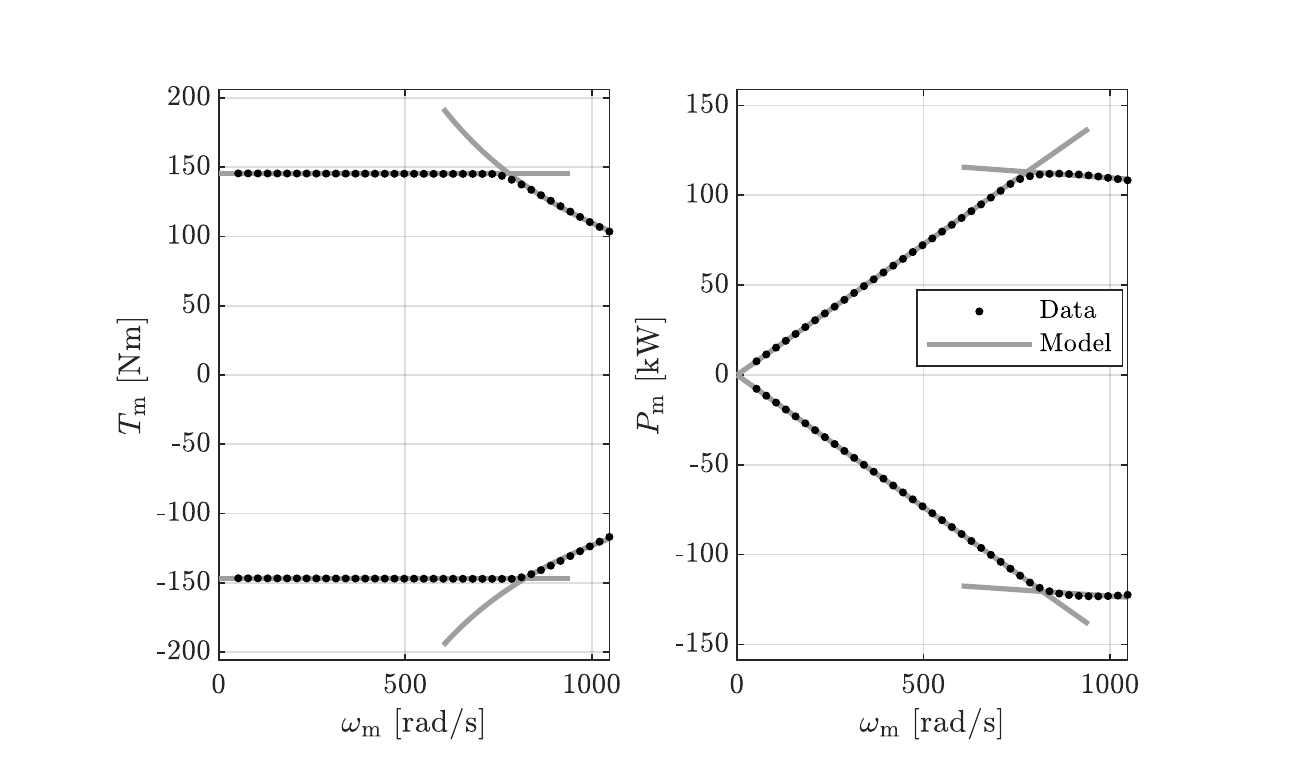}
	\caption{Identification of $T_\mathrm{m,min}$, $T_\mathrm{m,max}$, $d_\mathrm{0,min}$, $d_\mathrm{0,max}$, $d_\mathrm{1,min}$, $d_\mathrm{1,max}$ from MotorCAD data.}
	\label{fig:Motor_Limits}
\end{figure}

Fig.~\ref{fig:Motor_Limits} shows the identification of the constants $T_\mathrm{m,min}$, $T_\mathrm{m,max}$, $d_\mathrm{0,min}$, $d_\mathrm{0,max}$, $d_\mathrm{1,min}$, $d_\mathrm{1,max}$.

\subsection{Motor Loss Fitting} \label{app:Motor_Loss}
\begin{figure}[H]
	\centering
	\includegraphics[trim=1cm 0.45cm 1.75cm 0.75cm,clip=true, width=\columnwidth]{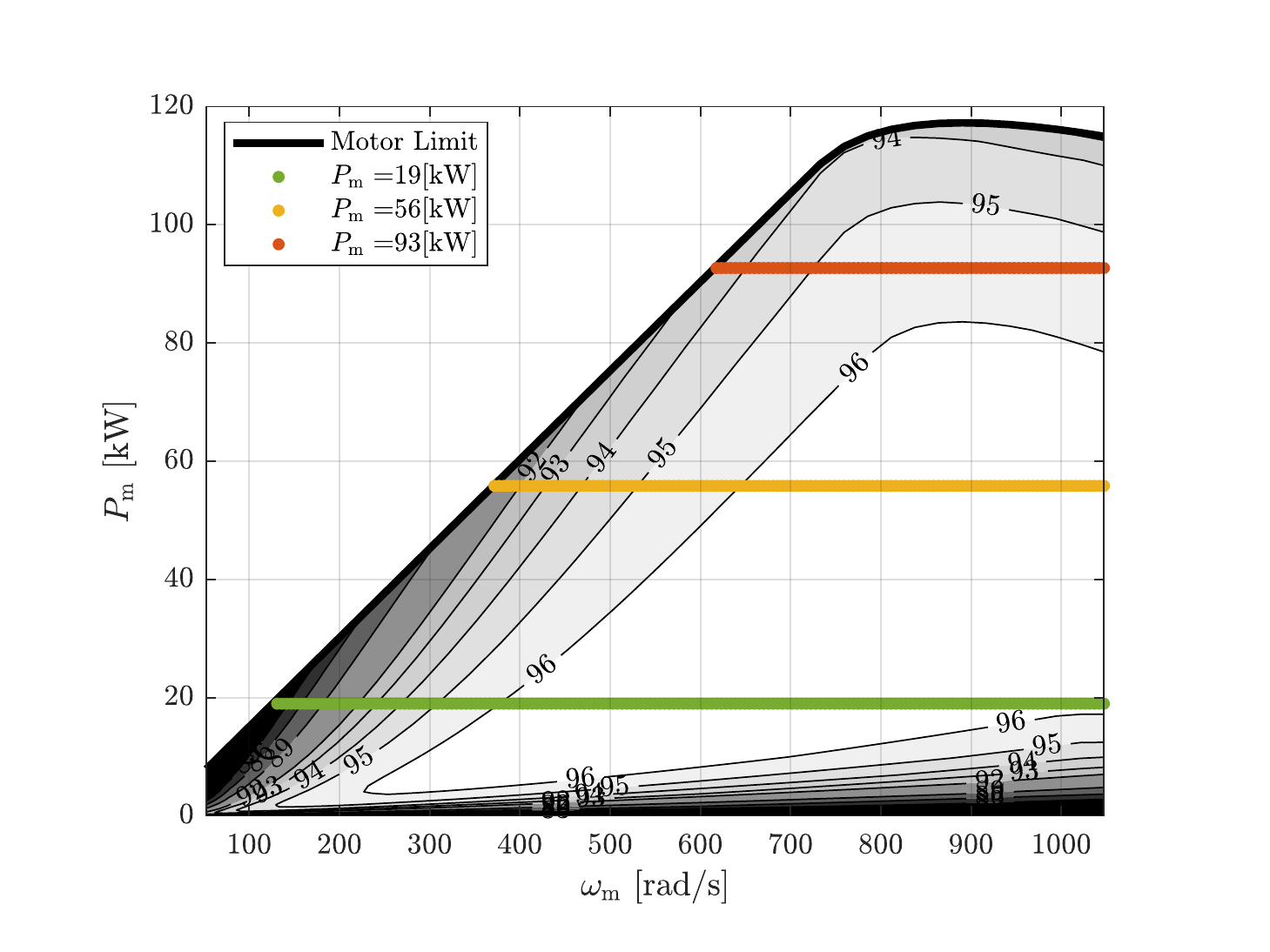}
	\caption{The power levels for state-of-the-art power loss model fitting.}
	\label{fig:Appendix_PowerLevels}
\end{figure}

\begin{figure}[H]
\centering
\includegraphics[trim=1.5cm 0.2cm 2.3cm 0.3cm,clip=true, width=\columnwidth]{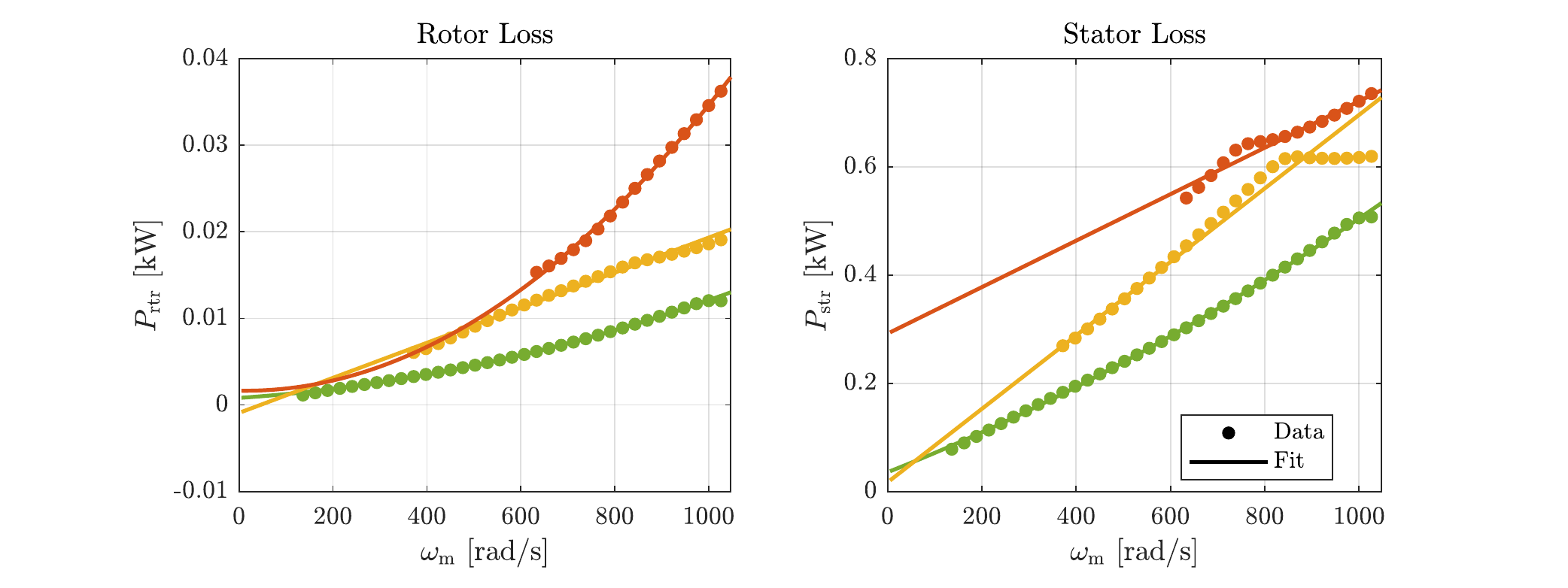}
\caption{Temperature-independent rotor losses (left) and stator losses (right) along with their models at power levels shown in Fig.~\ref{fig:Appendix_PowerLevels}.}
\label{fig:Appendix_PowerLevels_RtrStr}
\end{figure}

\begin{figure}[H]	
	\centering
	\includegraphics[trim=1.4cm 1.85cm 1.35cm 1.5cm,clip=true, width=\columnwidth]{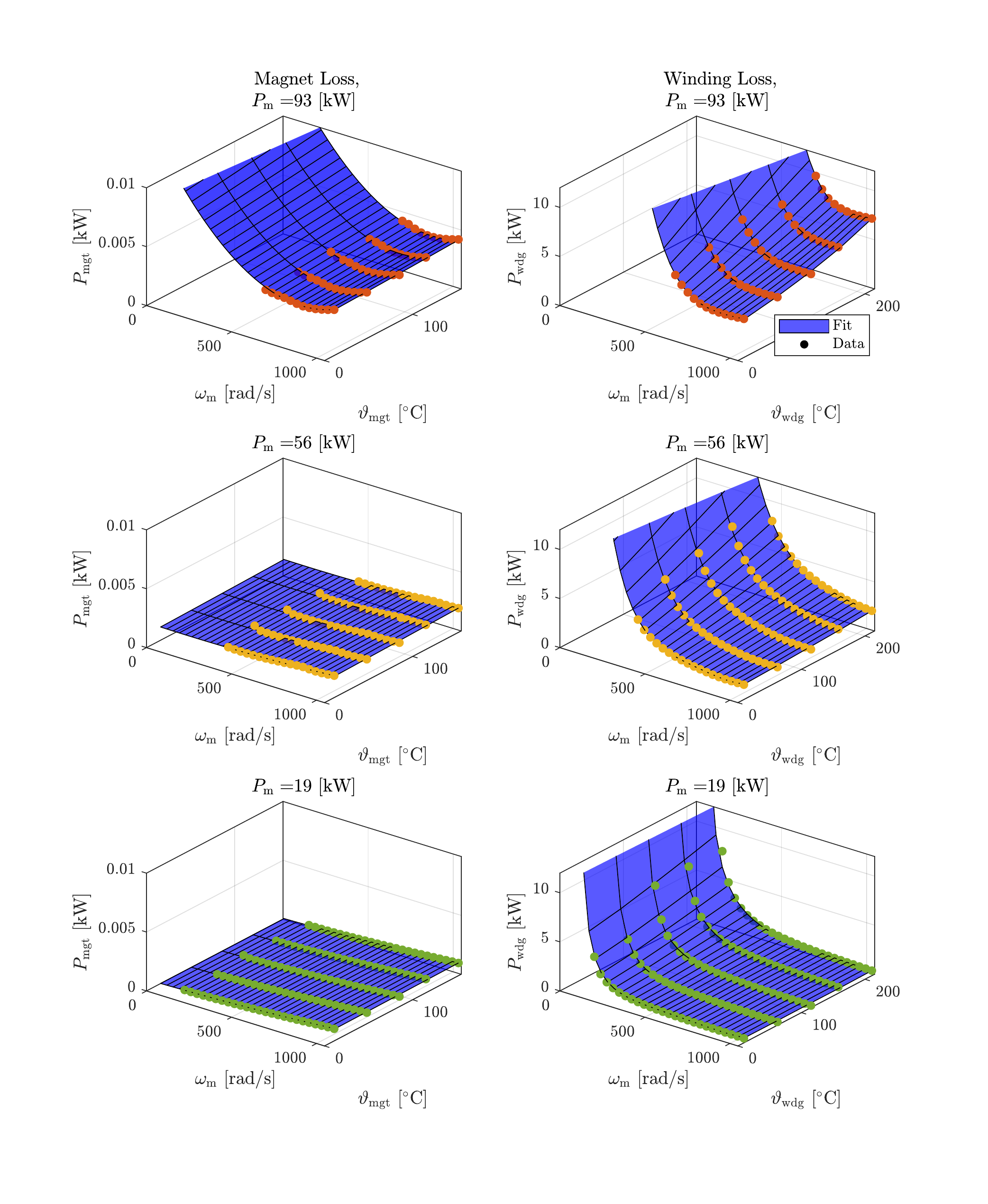}
	\caption{Temperature-dependent magnet losses (left) and winding losses (right) along with their models at power levels shown in Fig.~\ref{fig:Appendix_PowerLevels}.}
	\label{fig:Appendix_PowerLevels_MgtWdg}
\end{figure}

\subsection{Thermal Model Fitting} \label{app:Thermal_Fitting}
\begin{figure}[H]
	\centering
	\includegraphics[trim=1.25cm 0.75cm 1.35cm 0.75cm,clip=true, width=\columnwidth]{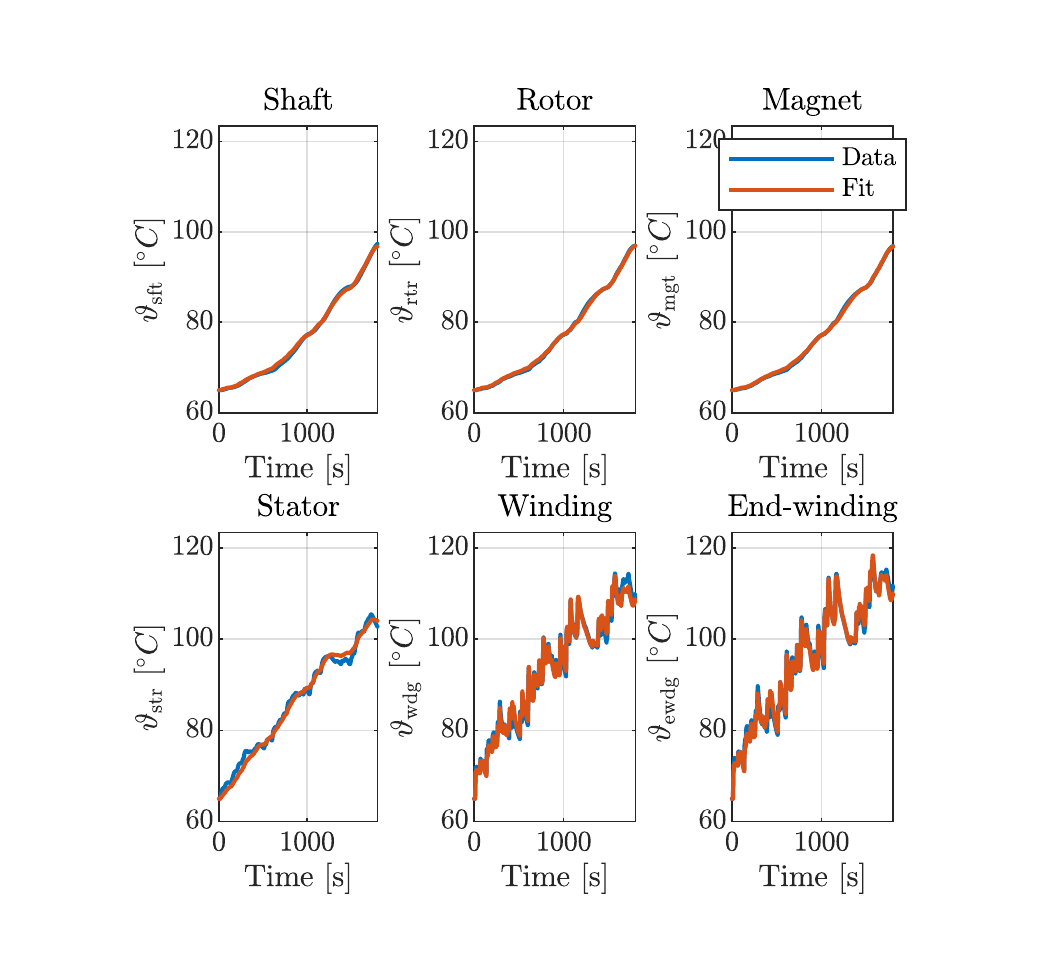}
	\caption{Identification of thermal coefficients. The collective NRMSE is 0.62\%.}
	\label{fig:Thermal_fit}
\end{figure}

\subsection{Inverter Identification} \label{app:Inveter_Loss}
\begin{figure}[H]	
	\centering
	\includegraphics[trim=1.25cm 0.3cm 1.7cm 0.6cm,clip=true, width=\columnwidth]{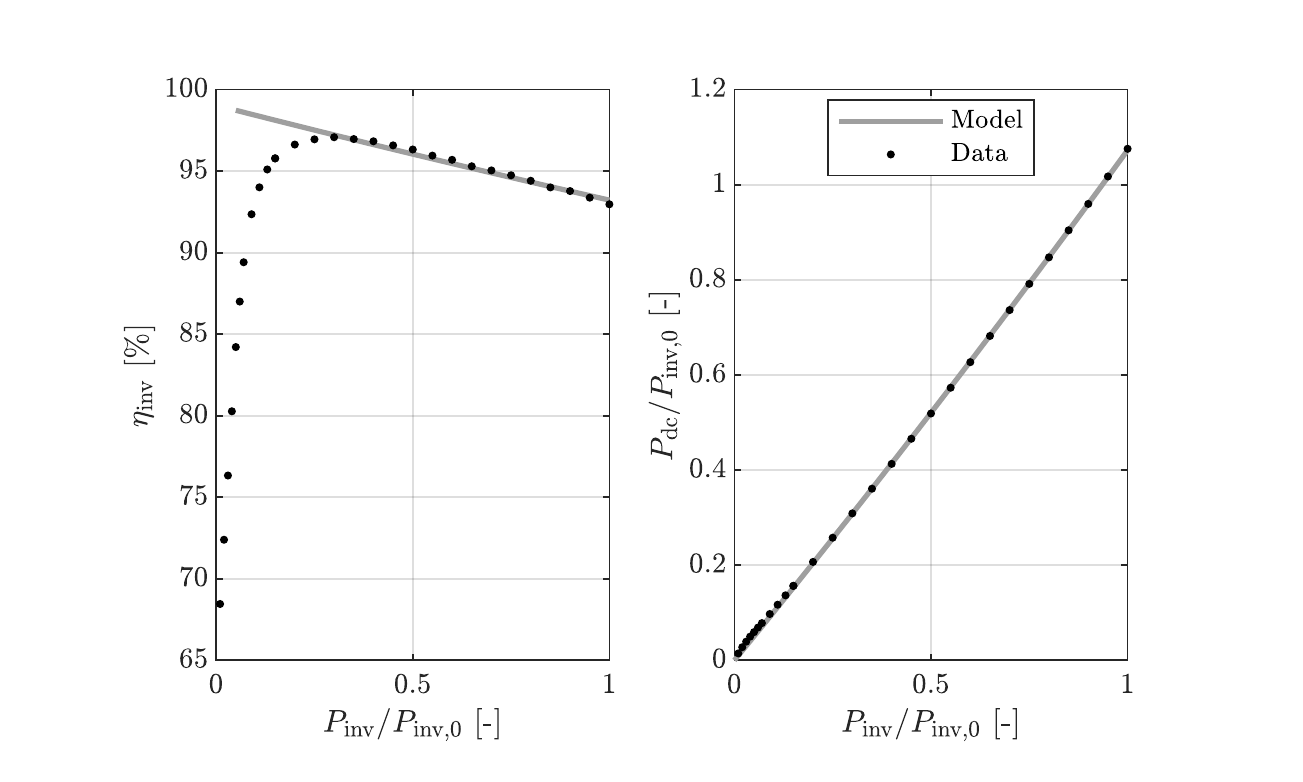}
	\caption{Identification of the inverter coefficient $\alpha_{\mathrm{inv}}$ with NRMSE$_{\mathrm{inv}} = 0.26\%$.}
	\label{fig:Inverter_Fit}
\end{figure}

Fig.~\ref{fig:Inverter_Fit} shows the identification of the constant $\alpha_{\mathrm{inv}}$.

\subsection{Radiator Fan Identification} \label{app:Fan_fitting}
\begin{figure}[H]
	\centering
	\includegraphics[trim=0.85cm 0.25cm 1.5cm 0.4cm,clip=true, width=\columnwidth]{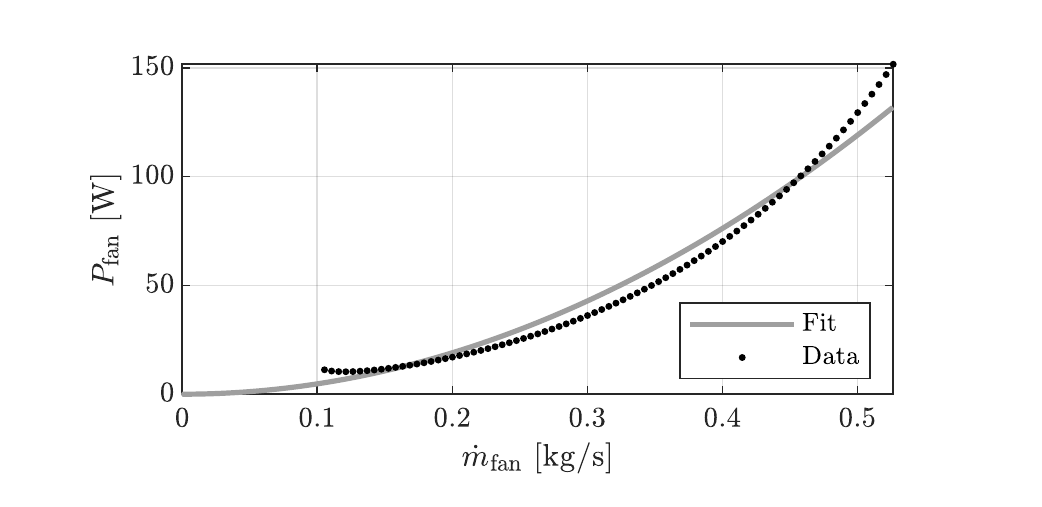}
	\caption{Identification of fan power coefficient $\alpha_{\mathrm{f}}$ with NRMSE$_{\mathrm{fan}} = 4.07\%$.} 
	\label{fig:Fan_fit}
\end{figure}%

\subsection{Battery Identification} \label{app:Battery_Fitting}
\begin{figure}[H]
	\centering
	\includegraphics[trim=0.65cm 0.3cm 1.45cm 0.5cm,clip=true, width=\columnwidth]{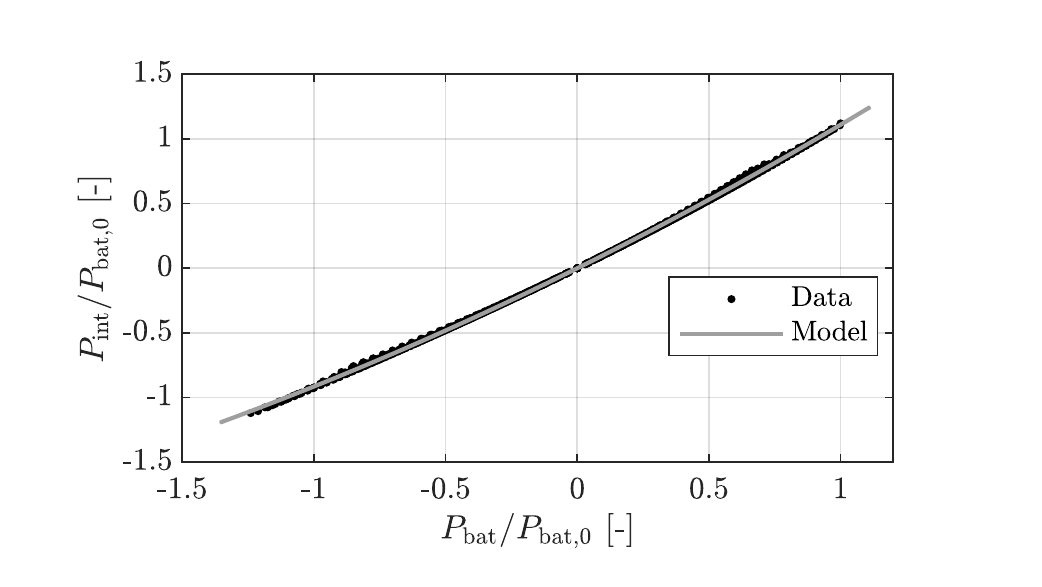}
	\caption{Identification of the battery coefficient $\alpha_\mathrm{b}$ with NRMSE$_{\mathrm{bat}} = 0.75\%$.}
	\label{fig:Appendix_Battery_quadratic_fit}
\end{figure}